\documentclass[useAMS,usenatbib]{mn2e}
\usepackage{bm}
\usepackage{pslatex}
\usepackage[pdftex]{graphicx,color}

\newcommand{\Psun}{\ensuremath{\,{\rm P}_{\odot}}}         % Period of Sun
\newcommand{\Pjup}{\ensuremath{\,{\rm P}_{\rm J}}}         % Period of Jupiter
\newcommand{\Rhill}{\ensuremath{\,{\rm R}_{\rm H}}}        % Hill radius symbol
\newcommand{\Msun}{\ensuremath{\,{\rm M}_\odot}}           % Solar mass symbol
\newcommand{\Rjup}{\ensuremath{\,{\rm R}_{\rm Jup}}}       % Jupiter radius symbol
\title{Application of the MEGNO technique to the dynamics of Jovian irregular 
satellites}

\author[Hinse et al.]
       {T. C. Hinse\,$^{1,3}$
        \thanks{E-mail: tobiash@astro.ku.dk},
        A.\ A.\ Christou\,$^{1}$,
	J.\ L.\ A.\ Alvarellos\,$^{2}$	
        \\
        $^1$\,Armagh Observatory, College Hill, Armagh, BT61 9DG, Northern 
	      Ireland, UK \\
        $^2$\,Space Systems/Loral MS G-76, 3825 Fabian Way, Palo Alto, CA
	      94303 \\
        $^3$\,Niels Bohr Institute, University of Copenhagen, Juliane Maries 
	Vej 30, Copenhagen \O, 2100, Denmark
        }

\begin{document} \maketitle %%%%%%%%%%%%%%%%%%%%%%%%%%%%%%%%%%%%%%%%%%%%%%%%%%%
%%%%%%%%%%%%%%%%%%%%%%%%%%%%%%%%%%%%%%%%%%%%%%%%%%%%%%%%%%%%%%%%%%%%%%%%%%%%%%%

\begin{abstract}
We apply the MEGNO (Mean Exponential Growth of Nearby Orbits) technique to the
dynamics of Jovian irregular satellites.  The MEGNO indicator is an effective
numerical tool to distinguish between  quasi-periodic and chaotic orbital time
evolution for a given dynamical system. We demonstrate the efficiency of
applying the MEGNO indicator to generate a mapping of relevant phase-space
regions occupied by observed jovian  irregular satellites. The construction of
MEGNO maps of the Jovian phase-space region within its Hill-sphere is
addressed and the obtained results are compared with previous studies
regarding the dynamical stability  of irregular satellites. Since this is the
first time the MEGNO technique is  applied to study the dynamics of irregular
satellites we provide a review of  the MEGNO theory. We consider the elliptic
restricted three-body problem in  which Jupiter is orbited by a massless test
satellite subject to solar gravitational perturbations. The equations of
motion of the system are  integrated numerically and the MEGNO indicator
computed from the systems  variational equations. An unprecedented large set
of initial conditions are  studied to generate the MEGNO maps. The chaotic
nature of initial conditions are demonstrated by studying a quasi-periodic
orbit and a chaotic orbit. As a result we establish the existence of several
high-order mean-motion resonances  detected for retrograde orbits along with
other interesting dynamical features. The computed MEGNO maps allows to
qualitatively differentiate between chaotic  and quasi-periodic regions of the
irregular satellite phase-space given only a  relatively short integration
time. By comparing with previous published  results we can establish a
correlation between chaotic regions and  corresponding regions of orbital 
instability.

\end{abstract}


%%%%%%%%%%%%%%%%%%%%%%%%%%%%%%%%%%%%%%%%%%%%%%%%%%%%%%%%%%%%%%%%%%%%%%%%%%%%%%%

\section{Introduction}
The existence of natural satellites in orbit around Solar System giant planets
has been known since Galileo discovered the four inner regular moons of 
Jupiter. Since then two classes of natural satellites have been identified to
orbit the outer giant planets: regular and irregular satellites. The study of 
irregular satellites is interesting as they are located close to the Hill 
sphere boundary of the parent planet. Consequently Solar perturbations are
present and  chaotic behaviour is expected in the time evolution of their
orbits. Recent  reviews on the subject of irregular satellites are given in 
\cite{Peale1999,JewittHaghighipour2007,Nicholson2008}. Regular satellites are 
characterised by circular orbits located close to the equatorial plane of the 
planet. This is the case of the Galilean satellites of Jupiter. The class of 
irregular satellites differ in being on large eccentric and highly inclined 
orbits in either prograde (direct) or retrograde motion.

The majority of the current population of giant planet irregular satellites
have been discovered by ground-based large aperture optical telescopes
operating in dedicated survey programs over the past 10 years
\citep{Gladman1998, Gladman2000, Gladman2001, SheppardJewitt2003, Holman2004,
Sheppard2005, Sheppard2006}. The most abundant irregular satellite population
is observed to exist at Jupiter counting approximately 60 members at the
current time. An interesting dynamical property of the phase space
distribution of orbital  elements is the clustering in distinct families for
both prograde and  retrograde irregular satellite members
\citep{nesvorny2003,nesvorny2004}.

The regular Galilean satellites are suggested to have formed by accretion
processes within a circumjovian planetary disk \citep{CanupWard2002, 
MosqueiraEstrada2003, EstradaMosqueira2006}. Observed kinematic differences
between regular and  irregular satellite orbits suggest a different formation
mechanism of irregular satellites. The most favoured formation scenario is
capture from an initial heliocentric orbit. The key element of permanent
capture is the necessity of a  frictional force (due to a gaseous environment) 
or energy transfer through close-encounter. Both processes are capable to 
dissipate orbital energy and provide a viable dynamical route to form a given
population of irregular satellites. For a description of various capture 
scenarios we refer to \citet{JewittHaghighipour2007}.

Several authors have studied general stability properties of irregular 
satellites. \cite{HaghighipourJewitt2008} studied the region between Callisto
and the innermost Jovian irregular satellite Themisto 
($30~\Rjup < r < 80~\Rjup$ where
$\Rjup$ is Jupiters radius and $r$ is the distance of the satellite to Jupiter). 
Despite observational evidence indicating that
this region is void of a satellite population they showed that a large 
fraction of this region renders possible satellite orbits stable for at least 
10 Myrs. Extensive stability surveys of irregular satellites close to the Hill 
sphere of the giant planets have been carried out by \cite{carruba2002, 
nesvorny2003} by recording the lifetimes of irregular satellite  test particles
in various parameter surveys. \cite{yokoyama2003} studied the  region
$250~\Rjup < r < 370~\Rjup$  for prograde Jovian irregular orbits in  the
semi-major axis and inclination plane finding evidence of the presence of
secular perturbations \citep{WhippleShelus1993, CukBurns2004, 
nesvornybeauge2007}. Stability properties of irregular 
orbits at the outer regions around the giant planet's Hill sphere and beyond 
were studied by \citet{ShenTremaine2008}.

Motivated to study the phase space topology structure of irregular satellites in
detail we applied the MEGNO chaos indicator \citep{cincottasimo2000,
gozdziewskiMEGNO,gozdziewskiMEGNO-2} to qualitatively differentiate
between quasi-periodic and chaotic  phase space regions. Initial tests showed
that MEGNO is efficient in showing chaotic regions using relative short
integration times of the orbit. In this work we outline the
basic principles of the MEGNO indicator as this is the first time this 
technique is applied to the dynamics of irregular satellites.

The structure of the paper is as follows. In section 2 we present the model and
numerical methods used as well as initial conditions and definitions of angular
variables. Section 3 presents a review of the MEGNO chaos indicator and an
outline of its computation. Its relation to the Lyapunov exponent is reviewed
and tests on numerical accuracy are presented. Section 4 describes the
construction of dynamical maps of Jovian irregular satellites as studied in 
this work. Section 5 presents and outlines essential steps to obtain the 
secular system of the time variation of a satellite's Keplerian elements using 
a time-averaged running window. Section 6 presents our results with comparison 
to previous work and section 7 concludes this work.

\section{Model, numerical methods and initial conditions}

The results obtained in this work are based on the elliptic restricted 
three-body problem. We integrate the equations of motion of the system
\citep{MorbidelliBook2002}

\begin{equation}
\frac{d^{2}\boldsymbol{r}_{i}}{dt^{2}} = -\frac{k^2 m_{0}}
{|\boldsymbol{r}_{i}|^{3}}~\boldsymbol{r}_{i}~+~\sum_{j=1, j \ne i}^{n} 
k^2 m_{j}\Bigg( \frac{\boldsymbol{r}_{j} - \boldsymbol{r}_{i}}
{|\boldsymbol{r}_{j} - \boldsymbol{r}_{i}|^{3}} - 
\frac{\boldsymbol{r}_{j}}{|\boldsymbol{r}_{j}|^3}\Bigg).
\label{EOM}
\end{equation}
\noindent
Here $n = 2$ in a jovian centric reference frame with $m_0$ denoting the mass 
of Jupiter and $k^2$ denotes the Gauss gravitational constant. The positions
(relative to Jupiter) and masses of the satellite and the Sun are 
$(\boldsymbol{r}_{1}, m_{1})$ and $(\boldsymbol{r}_{2},m_{2})$, respectively. In
this model the satellite is orbiting Jupiter and its orbit is subject to Solar
perturbations. In all integrations we regard the satellite as a test particle 
of zero mass. Oblateness effects from Jupiter are not considered in this 
model and perturbations from other planets are omitted as well. We consider this
simplified model for two reasons. First the purpose of this paper is to
demonstrate and apply the MEGNO chaos indicator to the dynamics of irregular
satellites and secondly we aim to qualitatively identify chaotic regions 
originating from Solar perturbations only. If we were to include oblatenes and 
planetary perturbations cause and effect would be very difficult to isolate. 
The orbit  of the satellite was obtained by the numerical integration of
Eq.~(\ref{EOM}) using  the 1) 15th-order Radau integration algorithm as
implemented in the latest version of the {\sc MERCURY} package
\citep{mercury6a,mercury6b} and 2) the Gragg-Bulirsch-Stoer (GBS) extrapolation
algorithm \citep{hairer1993} as implemented in the {\sc CS-MEGNO} code
\citep{gozdziewskiMEGNO}.

\begin{figure}
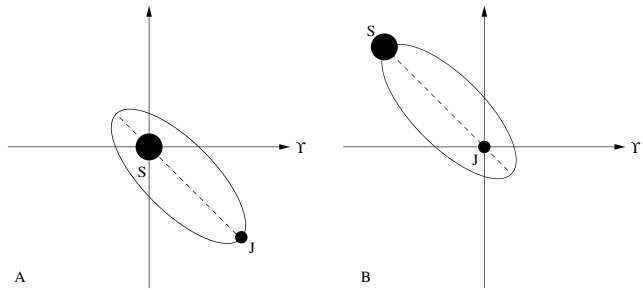

%Figure location on arpc2:
%/home/tobiash/phd-thesis/papers/letter
%XFIG name:HelioJovicentricRelationship.fig
%Note: The figure as created using XFIG. Using Export --> "EPS/Latex both 
%parts" generates *.pstex and *.pstex_t files. The *.pstex file was loaded into
%GIMP with 500 pix/in resolution and saved as a PNG file. The corresponding
%*.pstex_t was renamed to *.png_t and the include filename changeed.
\resizebox{83mm}{!}{\input HelioJovicentricRelationship.png_t}
%\scalebox{0.49}{\input{./HelioJovicentricRelationship.pdf_t}}
%\includegraphics{myfig.eps}
%\includegraphics[width=60mm]{myfig.eps}
%\includegraphics[height=60mm]{myfig.eps}
%\includegraphics[scale=0.75]{myfig.eps}
%\includegraphics[angle=45,width=52mm]{myfig.eps}
\caption{Graphical illustration of the relationship between angles in a
heliocentric reference system and a planetocentric system. In both panels
$\Upsilon$ denotes the reference direction from which inertial angles are
measured and the orbital motion is counterclockwise. \emph{Panel A}: Shows
the  orbit of (J)upiter with respect to the (S)un. Jupiters orbit is strongly
exaggerated  in its eccentricity for reasons of clarifications. In the
current situation the  longitude of pericenter of Jupiter is
$\varpi_{J}^{hel}=135^{\circ}$.  \emph{Panel B}: The orbit of (S)un with
respect to (J)upiter. Here the longitude of perijove $\varpi_{\odot}^{jov} =
315^{\circ}$.}
\label{HelioJovicentricRelationship}
\end{figure}

The Radau integrations were used to study individual  satellite orbits on the
long-time scales and the GBS integrator was used for  the generation of MEGNO
maps over the observed irregular satellite phase space  region. Initial
conditions (geometric cartesian elements relative to Jupiter's centre of mass) 
for the Sun and observed irregular satellites have been obtained from the JPL
Horizon \footnote{telnet ssd.jpl.nasa.gov 6775}  Ephemeris generator 
\citep{Giorgini1996} at the epoch 01-Jan-2005 (12:00 UT). 
It is worth to note that the Horizon Ephemeris generator cannot output
osculating  Keplerian elements of the Sun with respect to another object in the
Solar System. Only cartesian vectors can be retrieved for the Sun relative to
Jupiter. For the computations of the stability maps (see section 
\ref{stabilitymapsection}) we have transformed the Sun's cartesian elements to 
Keplerian elements using the {\sc mco\_x2el.f} subroutine as implemented in 
the latest version of {\sc MERCURY}. The transformation introduces 
round-off errors not larger than $10^{-10}$ in the absolute errors which is
much smaller than the error of observed elements of irregular satellites (e.g
15 meters in the semi-major  axis). The retrieved initial conditions are
referred to the ecliptic and mean  equinox  (xy-plane is the orbit of Earth)
at reference frame ICRF/J2000.0. In  this work we will present our results in
a \emph{planetocentric} reference system where Jupiter is at the centre and
the ecliptic is the reference  plane. Thus we denote the planetocentric
elements\footnote{the orbital  inclination of a satellite is measured relative
to the ecliptic.} of a  satellite as $(a,e,I,\omega,\Omega, M)$ (where $M$ is
the mean anomaly) and we  use the subscript $\odot$ to indicate the  Sun's
orbit relative to Jupiter and the subscript $J$ to denote the orbit of Jupiter
in the heliocentric reference system. For the longitude of perijove of
prograde irregular satellites we use the usual definition  $\varpi = \Omega +
\omega$. For retrograde satellites we use $\varpi =  \Omega - \omega$. The
mean longitude of an irregular satellite is then given  by $\lambda = \varpi +
M$.

In order to compare our data with previous published results in the literature 
using both planetocentric and heliocentric orbital elements we briefly point 
out the relationship of angles measured in the two reference frames. Changing 
the coordinate system from heliocentric to a jovicentric system leaves the
semi-major axis and eccentricity unchanged as these quantities are invariable
under a coordinate transformation. In Fig.~\ref{HelioJovicentricRelationship}
we demonstrate the relationship between the longitude of pericenter of the two 
bodies in the two reference system. If $\varpi_{J}^{hel}$ is the longitude of 
Jupiter in the heliocentric system and $\varpi_{\odot}^{jov}$ denotes the
longitude of the Sun in the jovicentric system, then from geometric arguments 
we have $\varpi_{J}^{hel} = \varpi_{\odot}^{jov} + 180^{\circ}$. A similar
argument will lead to the relationship  $\lambda_{J}^{hel} =
\lambda_{\odot}^{jov} + 180^{\circ}$ relating the mean longitudes of Jupiter 
and the Sun. The orbital inclination $(I_{\odot},I_{J})$ and the argument of 
node $(\Omega_{\odot}, \Omega_{J})$ are unchanged under the transformation.

\begin{figure*}
%Note: The original file is an EPS file. The EPS file was loaded into GIMP 
%with 500 pix/in resolution. The figure was then saved as a PNG file.
\centerline{
        \hbox{\includegraphics[width=87mm]{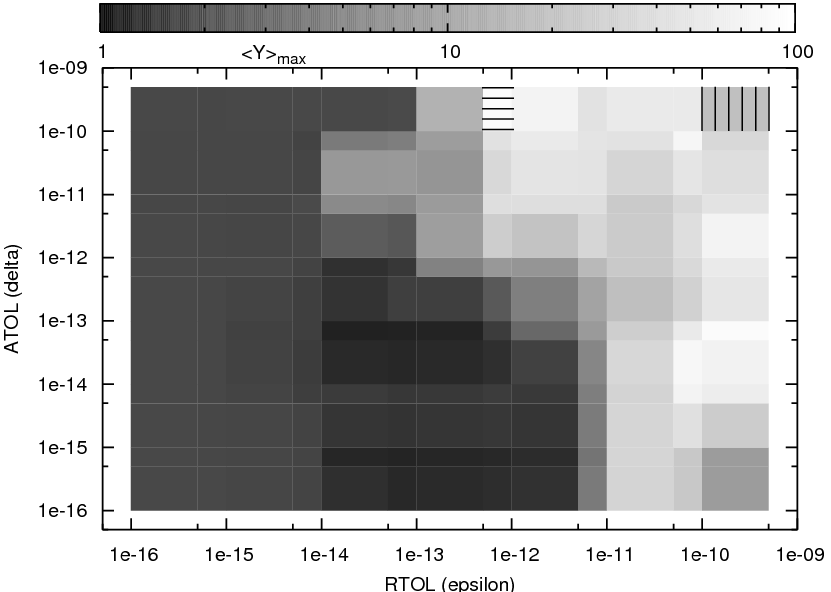}
              \includegraphics[width=87mm]{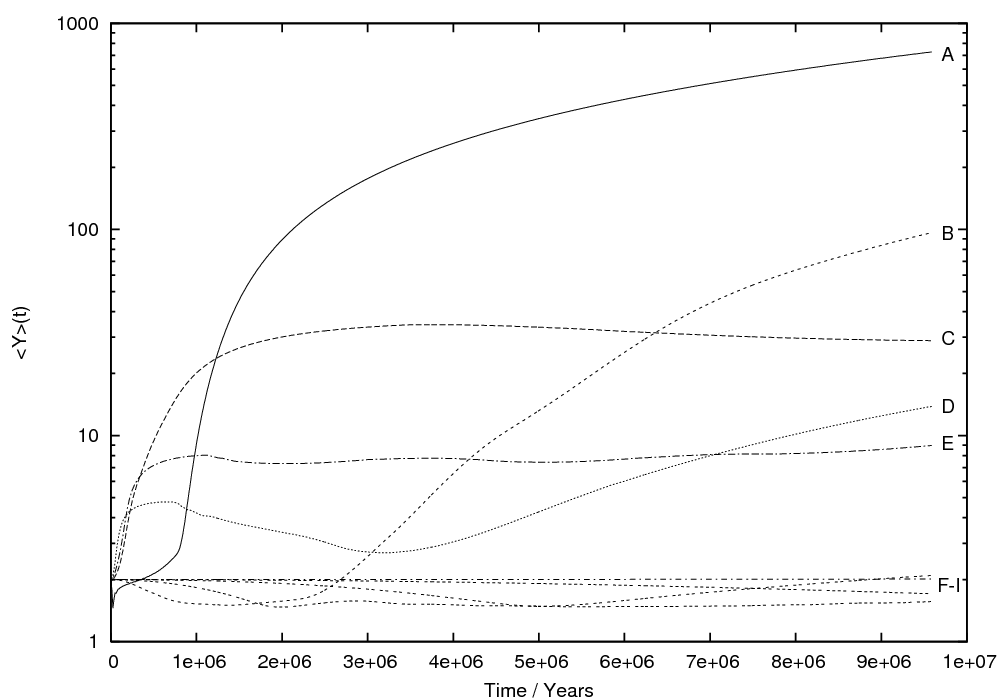}
             }
           }
\caption{Results from 2-body Kepler integrations. Right panel: 
A) $\textnormal{(RTOL,ATOL)}=(5.0\cdot10^{-11}, \cdot 10^{-10})$, 
B) $\textnormal{(RTOL,ATOL)}=(5.0\cdot10^{-13}, \cdot 10^{-10})$, 
C) $\textnormal{(RTOL,ATOL)}=(1.0\cdot10^{-11}, \cdot 10^{-10})$, 
D) $\textnormal{(RTOL,ATOL)}=(1.0\cdot10^{-10}, \cdot 10^{-10})$, 
E) $\textnormal{(RTOL,ATOL)}=(5.0\cdot10^{-11}, \cdot 10^{-10})$, 
F) $\textnormal{(RTOL,ATOL)}=(1.0\cdot10^{-16}, 1.0 \cdot 10^{-16})$, 
G) $\textnormal{(RTOL,ATOL)}=(1.0\cdot10^{-15}, 1.0 \cdot 10^{-15})$, 
H) $\textnormal{(RTOL,ATOL)}=(1.0\cdot10^{-14}, 1.0 \cdot 10^{-10})$, 
I) $\textnormal{(RTOL,ATOL)}=(5.0\cdot10^{-15}, 1.0 \cdot 10^{-12})$. 
The area marked with vertical lines corresponds to case D and horizontal line 
corresponds to case B. The final value after 10 million years of 
$\langle Y \rangle$ for case F) was 1.999757, case G) 2.000367, case H) 1.636787 and 
case I) 1.798234.}
\label{RTOLATOLMAP}
\end{figure*}

\section{The MEGNO chaos indicator}

The time evolution of irregular satellite orbits exhibits both quasi-periodic
and  chaotic dynamics \citep{WhippleShelus1993, SahaTremaine1993,
GoldreichRappaport2003}. An efficient numerical method to detect phase space
regions resulting in chaotic or quasi-periodic initial conditions is provided 
by the MEGNO (Mean Exponential Growth of Nearby Orbits) factor (or indicator).
The  MEGNO technique was first introduced by \citep{cincottasimo2000,
cincottasimo2003} and were originally inspired from the concept of
`conditional entropy of nearby orbits' (CENO) \citep{cincottasimo1999,
gurzadyan2000}. It can be applied to any dynamical system with more than two
degrees of freedom and has found widespread applications in dynamical
astronomy ranging from galactic dynamics to stability analysis of extrasolar
planetary systems and Solar System small body dynamics \citep{cincottasimo2003,
gozdziewskiMEGNO,gozdziewski2002, gozdziewski2004, breiter2005}.

By numerically evaluating the MEGNO factor after a given integration time one
obtains a quantitative measure of the degree of stochasticity of the system.
One method \citep[and references therein]{MorbidelliBook2002,DvorakBook2005}
to discriminate between ordered (or regular) and chaotic (very often 
associated with orbital instabilities) satellite orbits, is the calculation of
the system's Lyapunov characteristic exponents (LCE) or Lyapunov characteristic
numbers (LCN). From the theory of dynamical systems each system has a spectrum
of Lyapunov exponents (possibly complex eigenvalues) each associated to a given
 eigenvector of the system. The Lyapunov exponents describe the rate of change
of its corresponding eigenvector in time. The number of positive (or vanishing)
Lyapunov exponents indicates the number of independent directions in phase
space along which the satellite orbit exhibits chaotic (or quasi-periodic)
behavior \citep{MorbidelliBook2002}. In particular, MEGNO is closely related to
 the maximum Lyapunov exponent (MLE or sometimes referred to maximum Lyapunov
numbers, MLN) providing a quantitative measure of the exponential divergence of
nearby orbits and belongs to the class of fast Lyapunov indicators 
\citep{MorbidelliBook2002}.

In this work, we apply the MEGNO criterion to the dynamics of Jovian irregular
satellites. Details on the MEGNO concept and its numerical  computation can be
found in \cite{cincottasimo2003} and \cite{gozdziewskiMEGNO}. Since the  MEGNO
technique is applied for the first time to the dynamics of irregular
satellites,  we give a short review of the most important aspects of MEGNO and
provide additional information on its numerical computation.

\subsection{The Lyapunov Exponent}

The maximum Lyapunov exponent $\gamma$, provides a useful quantitative measure 
to study the dynamical nature of the time evolution of a satellite orbit in 
phase space and is defined as
\begin{equation}
\gamma := \lim_{t \rightarrow \infty} \frac{1}{t-t_{0}}
\ln\Bigg(\frac{\delta(t)}{\delta(t_{0})}\Bigg) = 
\lim_{t \rightarrow \infty}\frac{1}{t-t_{0}} \int_{0}^{t}
\frac{\dot{\delta}(s)}{\delta(s)}ds,
\end{equation}
\noindent
where $\delta(t)$ is the variational vector and measures the distance in phase 
space between two initially nearby orbits as a function of time. The second 
equality is easily realised by change of variables\footnote{If $y = \delta(s)$,
then $\int_{0}^{t} (\dot{\delta}(s)/\delta(s))ds = \int_{0}^{t}(1/y)dy = \ln
(y(t)/y(0))$}. For $\gamma > 0$, an initial separation grows exponentially in
time (definition of chaotic motion) at the rate $e^{\gamma t}$ or decays, if
$\gamma < 0$. In the case $\gamma = 0$, the time rate of change of the
variational vector $\dot{\delta}$  is 0. Since the (elliptic) restricted 
three-body problem is a conservative system the case $\gamma < 0$ is never 
encountered and would indicate the presence of dissipative forces. In 
practical computations only a finite time estimate $\gamma(t)$ of $\gamma(s)$ is
obtained after integration time $t$. In our computations, we will pay some
attention on the proper choice of $t$. Usually the convergence of $\gamma$(t)
is slow and a reliable numerical estimate of $\gamma$ requires a long
integration time of the dynamical system. When exploring the dynamics of a
large portion of phase  space, short integration times are preferred when
exploring the phase-space structure. For classic computations of the Lyapunov 
exponents from the variational equations we refer to \citep{Benettin1980,
Wolf1985}.

\subsection{Mathematical properties of MEGNO}

A faster convergence property is obtained from the mean exponential growth
factor of nearby orbits. The MEGNO indicator is closely related to the
definition of $\gamma$ and is defined as
\begin{equation}
Y(t) = \frac{2}{t}\int_0^t\frac{\dot{\delta}(s)}{\delta(s)}~s~ds,
\label{MEGNO}
\end{equation}
along with its time-averaged mean value
\begin{equation}
\langle Y\rangle(t) = \frac{1}{t}\int_0^tY(s)~ds.
\label{time-averageMEGNO}
\end{equation}
\noindent
\cite{cincottasimo2000} showed that $\langle Y \rangle(t)$ converges faster to
its limit value than the Lyapunov characteristic number. In the former
definition the relative rate of change of the separation vector 
$\dot{\delta}/\delta$, is weighted with time during the integration giving 
preference to the memory of late evolutionary behavior of the separation 
vector \citep{MorbidelliBook2002}. The time-weighting factor introduces an
amplification of any stochastic behavior, allowing an early detection of 
chaotic motion \citep{gozdziewskiMEGNO}.

Computing the time evolution of Eq.~(\ref{time-averageMEGNO}) for a set of initial
conditions allows the study of the dynamical properties in a given phase space 
region. Following \cite{cincottasimo2000, cincottasimo2003}, if the motion is
of quasi-periodic  nature, then $\delta$ grows linearly in time and $\langle
Y\rangle$ will exhibit asymptotic oscillations about 2 for $t \rightarrow
\infty$. In the case of chaotic initial conditions, $\langle Y \rangle(t) =
\gamma t/2$ for  $t \rightarrow \infty$. The latter equation relates the
Lyapunov characteristic number to the limiting value of $\langle Y \rangle(t)$
at integration time $t$. A linear best fit to $\langle Y \rangle(t)$ would
recover the Lyapunov characteristic number \citep{gozdziewskiMEGNO} . In 
summary, the time-averaged 
MEGNO ($\langle Y \rangle(t)$) converges faster to its limit value compared to the
standard calculation of the Lyapunov characteristic number. This property allows
a more rapid exploration of the dynamical phase space structure.

\subsection{Numerical calculation of $Y(t)$ and $\langle Y\rangle(t)$}
In practice, $Y(t)$ and $\langle Y\rangle(t)$ are calculated by rewriting
Eqs.~(\ref{MEGNO}) and (\ref{time-averageMEGNO}) into two differential 
equations \citep{gozdziewskiMEGNO}
\begin{equation}
\frac{dx}{dt} = \frac{\dot{\boldsymbol{\delta}}}{\boldsymbol{\delta}}~t~~~~~
\textnormal{and}~~~~~\frac{dw}{dt} = 2~\frac{x}{t},
\end{equation}
\noindent
where $Y(t) = 2 x(t)/t$ and $\langle Y\rangle(t)=w(t)/t$. To obtain 
$x(t),w(t)$ the two first-order coupled differential equations are solved 
along with the equations of motion. Initial condition for the
variationals are chosen using a random generator. The quantities 
$\delta = \int\delta 
\dot{r}$ and $\dot{\delta} = \int\delta \dot{v}$ of the $i$'th body are 
obtained by solving the variational equations $\boldsymbol{\delta}=
\delta\boldsymbol{v}$ and $\dot{\boldsymbol{\delta}}=
\delta{\boldsymbol{\dot{v}}}$ \citep{MikkolaInnanen1999}. At each time step we
have for the $i-$th body the variationals

\begin{eqnarray}
\delta \dot{r}_{i}&=&\delta v_{i} \label{delta} \label{delta}\\
\delta \dot{v}_{i}&=&-k^2(m_{0}+m_{i})\Bigg( 
\frac{\delta\boldsymbol{r}_{i}}{\vert \boldsymbol{r}_{i}\vert^3} - 
\frac{3\delta\boldsymbol{r}_{i}(\boldsymbol{r}_{i} \cdot 
\delta\boldsymbol{r}_{i})}{\vert \boldsymbol{r}_{i}\vert^5}\Bigg) - \delta
\boldsymbol{A}_{i} \label{delta-dot},
\end{eqnarray}
\noindent
the first term describes the variation in the 2-body Kepler orbit and the 
variation in perturbations (or interactions) is
\begin{eqnarray}
\delta \boldsymbol{A}_{i} &=& - \sum_{j\ne 0}^{i}m_{j}
\Bigg\{\Bigg[ \frac{\delta \boldsymbol{r}_{ij}}{\vert\boldsymbol{r}_{ij}
\vert^3} - \frac{3\boldsymbol{r}_{ij}(\boldsymbol{r}_{ij}\cdot\delta
\boldsymbol{r}_{ij})}{\vert \boldsymbol{r}_{ij}\vert^5}     \Bigg ] \\ 
 & & - \Bigg[\frac{\delta \boldsymbol{r}_{0j}}{\vert\boldsymbol{r}_{0j}
\vert^3} - \frac{3\boldsymbol{r}_{0j}(\boldsymbol{r}_{0j}\cdot\delta
\boldsymbol{r}_{0j})}{\vert \boldsymbol{r}_{0j}\vert^5} \Bigg ]\Bigg\}.
\label{deltaA}
\end{eqnarray}
\noindent
Here, $\boldsymbol{r}_{ij} = \boldsymbol{r}_{j} - \boldsymbol{r}_i$,
$\delta\boldsymbol{r}_{ij} = \delta\boldsymbol{r}_{j} - \delta\boldsymbol{r}_i$
and $\delta\boldsymbol{r}_{0j} = \delta\boldsymbol{r}_{j}-
\delta\boldsymbol{r}_{0}$, where the zero superscript denotes the central body.
Eqs.~(\ref{delta})-(\ref{delta-dot}) are computed in a straightforward way 
once the initial conditions have been defined for the variationals and the
initial osculating orbits. 

\subsection{Numerical accuracy tests}
The MEGNO indicator is numerically computed as outlined in the previous
section using the Gragg-Bulirsch-Stoer integration algorithm
\citep{hairer1993}. All computations are carried out using double precision
arithmetic. To control the numerical  errors during computations two accuracy
parameters are necessary for a GBS  integration. The two parameters control
the absolute $(\delta)$ and relative  $(\epsilon)$ error tolerances for any
given integration and both needs to be  specified for a given accuracy
requirement. In practical computations the usual choice of
$(\delta,\epsilon)$ falls into the range from one part in $10^{9}$ down to
the limit of the machine precision $(10^{-16})$ (depending on 
architecture). To find a suitable $(\delta,\epsilon)$ set, we integrated
several 2-body Kepler problems for 10 million years by following the orbit of
Himalia.  In that case the motion is expected to be quasi-periodic, and we
use it as a  test case to determine confidence of the computed MEGNO
indicator which should  converge to $\langle Y\rangle(t) = 2.0$. Initial
numerical tests showed that the  convergence of $\langle Y\rangle$ depends on
the accuracy of the numerical  integration. This property was already
outlined in \cite{gozdziewskiMEGNO}. We integrated Himalia's orbit (without
Solar perturbations) for different  combinations of the tolerance parameters.
In total we considered $(14 \times 14)$ combinations with $(\delta,\epsilon)$
 ranging from $10^{-10}$ to $10^{-16}$. The result of the integrations is
shown  in Fig.~\ref{RTOLATOLMAP}. The left figure panel shows the numerical
value  of $\langle Y \rangle$ (gray-scale color coded) at the end of the
integration for a given combination of $(\delta,\epsilon)$. The right panel
in Fig.~\ref{RTOLATOLMAP} shows the time evolution of $\langle Y\rangle(t)$
of a  few selected combinations of the tolerance parameters. The
$(\delta,\epsilon)$ combinations resulting in the cases denoted by (A-E) are
evidently of poor  numerical accuracy showing $\langle Y \rangle \gg 2.0$
after 10 million years. We interpret this result as artificial numerical
chaos, the result of the poor resolution of time descretisation in the
integration algorithm. The cases  (F,G,H,I) show a significant better
agreement with the expected value of  $\langle Y\rangle = 2.0$ with some
numerical fluctuation of $\langle  Y\rangle$ observed during the integration.
From the numerical experiments we fix
$(\delta,\epsilon)=(10^{-16},10^{-15})$ for \emph{all} MEGNO computations in
the present work. In addition, we monitor the relative energy error $dE/E$ and
note the maximum value reached during a given integration. For our choice of 
the tolerance parameters the average maximum relative energy error is on the
order of $dE/E \sim 10^{-11}$ over 1 million years. By examining several spot
tests on the time evolution of the relative energy error no systematic trends
were observed only exhibiting a random walk over time. For all orbits the 
maximum relative energy error were smaller than $10^{-12}$ though in the
restricted three-body  problem this does not reflect the accuracy of the
orbit of a massless test  satellite. For that reason we follow the same
approach as outlined in \cite{nesvorny2003} and monitored the Jacobi constant
during the numerical integration and found a maximum relative change of this
quantity to be on the order of typically $10^{-6}$. We also computed the 
absolute error of the semi-major axis ($\vert
a_{\odot}(t)-a_{\odot} \vert$)  of the Sun and found it to be less than a few
parts in $10^{-12}$. This means a preservation of 8 significant digits of the
semi-major axis over 60000 years. Since we are not aiming at generating
high-accuracy ephemerides of irregular satellite orbits we consider these
tests as sufficiently reliable in order to establish confidence in our
results presented in this work. In addition, uncertainties in observed
orbital elements of irregular satellites are assumed to be much greater than
rounding/truncation errors introduced by the numerical  integration
algorithm.

\begin{figure}
\includegraphics[width=8.5cm, height=6cm]{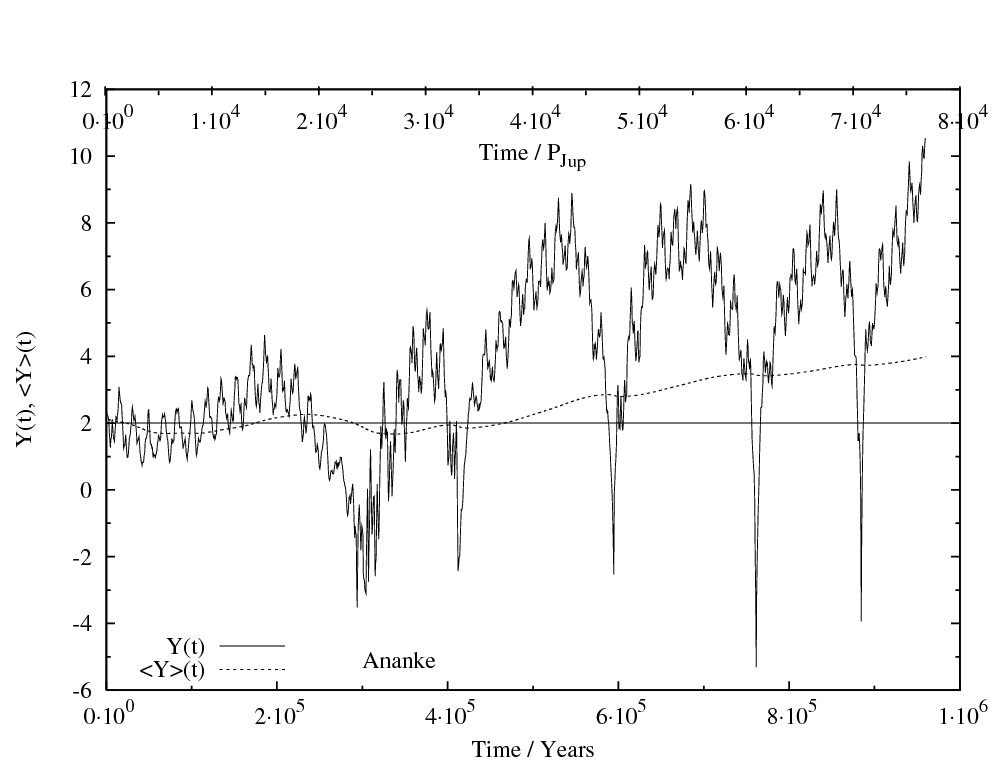}
%Note: The original file is an EPS file. The EPS file was loaded into GIMP 
%with 500 pix/in resolution. The figure was then saved as a PNG.
\caption{Time evolution of $Y(t)$ (solid curve) and $\langle Y\rangle(t)$ 
(dotted curve) from a numerical integration over 1 Myrs for the 
retrograde satellite Ananke including Solar perturbations. The solid 
horizontal line indicates $\langle Y \rangle = 2.0$. The upper time line 
expresses integration time in units of Jupiters orbital period $\Pjup \approx
12.5$ yrs.}
\label{anankemegno}
\end{figure}

\begin{figure}
%Note: The original file is an EPS file. The EPS file was loaded into GIMP 
%with 500 pix/in resolution. The figure was then saved as a PNG.
\includegraphics[width=8.5cm, height=6cm]{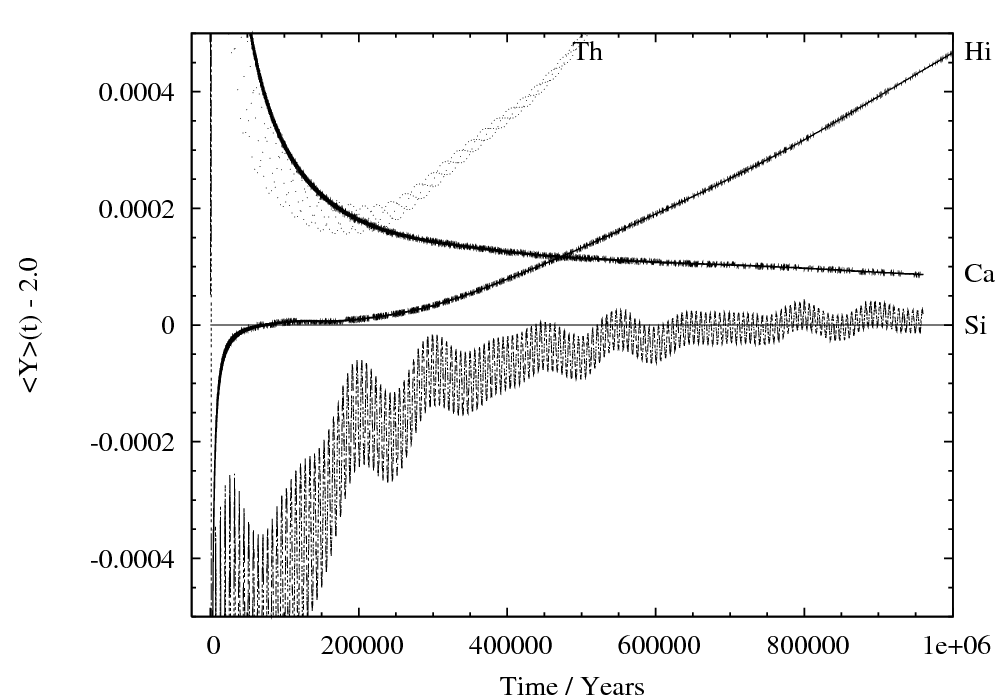}
\caption{Time evolution of $\langle Y\rangle$ for different selected irregular
satellites over a time scale of 1 million years. The shown satellites are 
Th(emisto), Hi(malia), Ca(rme) and Si(nope).}
\label{Y.DifferentSatellites}
\end{figure}

\subsection{Numerical examples of test orbits and limitations}
As an example Fig.~\ref{anankemegno} shows the time
evolution of $Y(t),\langle Y \rangle(t)$ for the irregular satellite Ananke
over a time span of 1 Myrs. In this integration, gravitational
perturbations from the Sun were included and it is observed that Ananke's
orbit exhibits a weak sign of chaoticity. It is evident that no clear 
convergence of $\langle Y\rangle \rightarrow 2$ is observed. Also the time
evolution of $Y(t),\langle Y \rangle(t)$ suggests that the time-averaged 
MEGNO $(\langle Y \rangle)$ is more reliable to indicate the presence of 
chaotic behavior as $Y(t)$ exhibits large variations about $Y(t)=2.0$ during 
the integration. 

In Fig.~\ref{Y.DifferentSatellites} we show the results of calculating $\langle
Y \rangle$ of four selected irregular satellites (Carme, Himalia, Sinope and
Themisto) over 1 Myr. Our results suggest that both Themisto and Himalia (both
on prograde orbits) shows chaotic behaviour as their calculated  $\langle Y
\rangle$ show clear sign of divergence from 2.0. The contrary is  seen for the
orbits of Carme and Sinope (both on retrograde orbits)  indicating convergence
(although slow for Carme) towards 2.0 indicating  quasi-periodic time
evolution. This result is in agreement with what is  expected from numerical
simulations. Prograde orbits tend to be less stable compared to retrograde
orbits  \citep{HamiltonKrivov1997,nesvorny2003}. It is noteworthy that although
 the calculation of $\langle Y \rangle$ is a powerful numerical tool for the
detection of chaotic initial conditions it is necessary to be cautious when
interpreting results. As mentioned previously two of the irregular satellites
(Themisto and Himalia) show a weak sign of chaotic  behaviour with $\langle Y
\rangle$ diverging from 2.0 with  $|\langle Y \rangle - 2.0| < 0.0003$ after
200000 years. In a MEGNO  map such an initial condition would appear as
quasi-periodic considered the large range in $\langle Y \rangle$ covering
several orders of magnitude. Therefore it is imperative to point out that every
numerical tool used to differentiate between quasi-periodic and chaotic
behaviour is only capable of showing quasi-periodicity up to the integration
time. On the contrary once chaotic behaviour (excluding numerical chaos) has
been detected the orbit can be claimed chaotic for all times. If each initial
condition in the MEGNO maps  presented in this work were calculated on time
scales similar to the age of the Solar System it is likely that no
quasi-periodic orbits are detected. In our maps we can with high confidence
associate chaotic regions with unstable orbits where the satellite either
escapes or experience a collision with one of the inner moons or if the
eccentricity grows high enough (due to the Kozai mechanism) the satellite may
even collide with Jupiter itself.

\section{Construction of MEGNO, ME, MS and MI maps}
In this work we color code the MEGNO indicator on a 2-dimensional phase space 
section mapping out the dynamics of irregular satellites. In particular we show
the dynamics in $(a,I)-$space, where $a$ is the semi-major axis and $I$ denotes
the orbit inclination. We chose initial conditions in which orbit inclinations
are relative to ecliptic (see IC section). For a given range in semi-major axis 
and inclination the grid of initial conditions in the maps are given by
\begin{figure*}
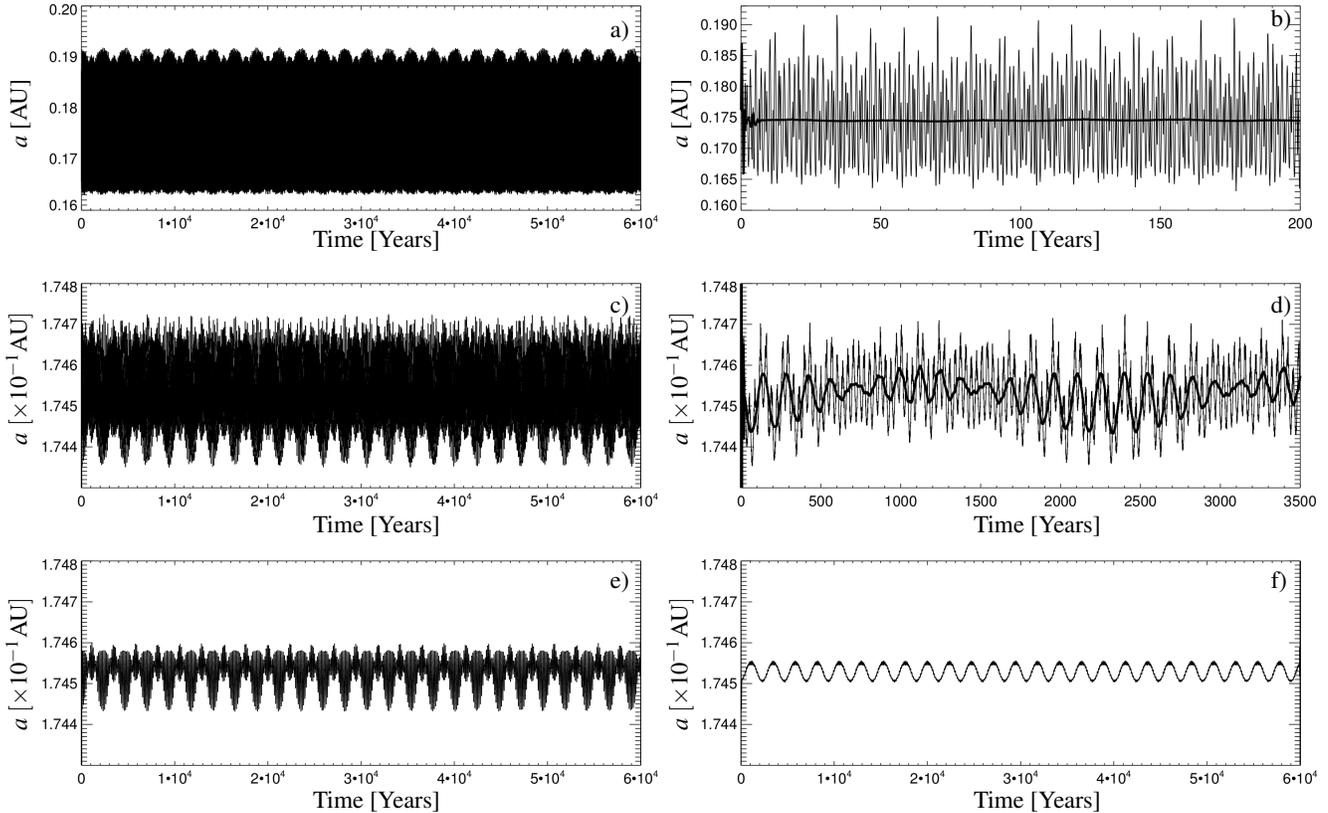

\centering
%Figure information:
%Path:/home/tobiash/simulations/jupiter/SingleOrbitIntegrationsForTobias/
%MeanMotionResonances/5:1MMR/Orbit5Quasiperiodic
%IDL script:DemonstratingSmoothing.pro
%Comments: Figure has been edited by using XFIG in order to add the axis labels.
\resizebox{175mm}{!}{\input DemonstratingSmoothing.png_t}
\caption{Demonstrating the process of successive smoothing using a running 
window time average applied on the semi-major axis. See text for details.}
\label{SmoothingSemiMajorAxis}
\end{figure*}
\begin{equation}
a_{i} = a_{min} + \frac{\Delta a}{N_{x}}~i = a_{min} + 
\frac{a_{max} - a_{min}}{N_{x}}~i
\end{equation}
\noindent
and
\begin{equation}
I_{j} = I_{min} + \frac{\Delta I}{N_{y}}~j = I_{min} + 
\frac{I_{max} - I_{min}}{N_{y}}~j,
\end{equation}
\noindent
where $i = 0, \ldots N_{x}$ and $j = 0, \ldots N_{y}$ are integers defining
the grid resolution. Chosing a large $(N_{x},N_{y})$ will result in a more
detailed mapping of phase-space structures in $(a,I)$ space. The range in
semi-major axis is chosen to span $a \in [0.04,0.20]$ AU 
($[0.11,0.56] \Rhill, [83,419] \Rjup$). Satellite inclinations cover the 
range $I \in [0,180]$ degrees considering  both prograde and retrograde
satellite orbits. In all MEGNO maps we chose to color code initial conditions
resulting in quasi-periodic motion by blue (see the electronic version of
this work). In the quasi-periodic case we have $\vert \langle Y\rangle -
2.0\vert < 0.01$ at the end of the numerical integration. In addition, we
also provide maps showing the  maximum eccentricity (ME) of an orbit at a
given initial grid point. ME maps  were constructed in parallel with the
MEGNO calculations. The osculating  elements were measured at each completed
Jovian period and the maximum value  is determined by comparing with a
previous measurement of the eccentricity.

\section{Smoothing orbital elements}

To study the secular system of irregular satellites we have to filter out the 
fast frequencies. \cite{SahaTremaine1993} already pointed out that the
unfiltered variations in the action elements (semi-major axis, eccentricity
and inclination) are larger than the filtered time series of a given 
element. This means that the fast variations (high-frequency terms) are much 
larger in amplitude than the slow variations (long-period terms). Hence the 
secular system is masked by the high frequencies. In this work an initial 
preliminary study of several test orbits confirmed this dynamical behaviour 
and we find that the most interesting dynamical features are to be found in 
the secular system. To obtain  the secular system one can either average out
the fast frequencies by applying  a digital filter in either the time or
frequency domain 
(\cite{Carpino1987,Quinn1991,SahaTremaine1993,Michtchenko1995}) or by simply
averaging out all quasi-periodic oscillations with a running window average
\citep{Morbidelli1997,MorbidelliNesvorny1999}.
    
In order to study the secular system we smooth the orbital elements of a given
time sequence  $A_{i}$ using the {\sc SMOOTH} function as implemented in
{\sc IDL}\footnote{\texttt{IDL} stands for Interactive Data Language. For
more information \texttt{http://www.ittvis.com/ProductServices/IDL.aspx}}. The
smoothing procedure is a running window average applied on  the full data set
as obtained from a numerical integration. The output of a given numerical
integration is given as a time sequence of orbital elements sampled at regular
intervals of length $\Delta T$. For a given sequence of an  orbital element
\begin{figure*}
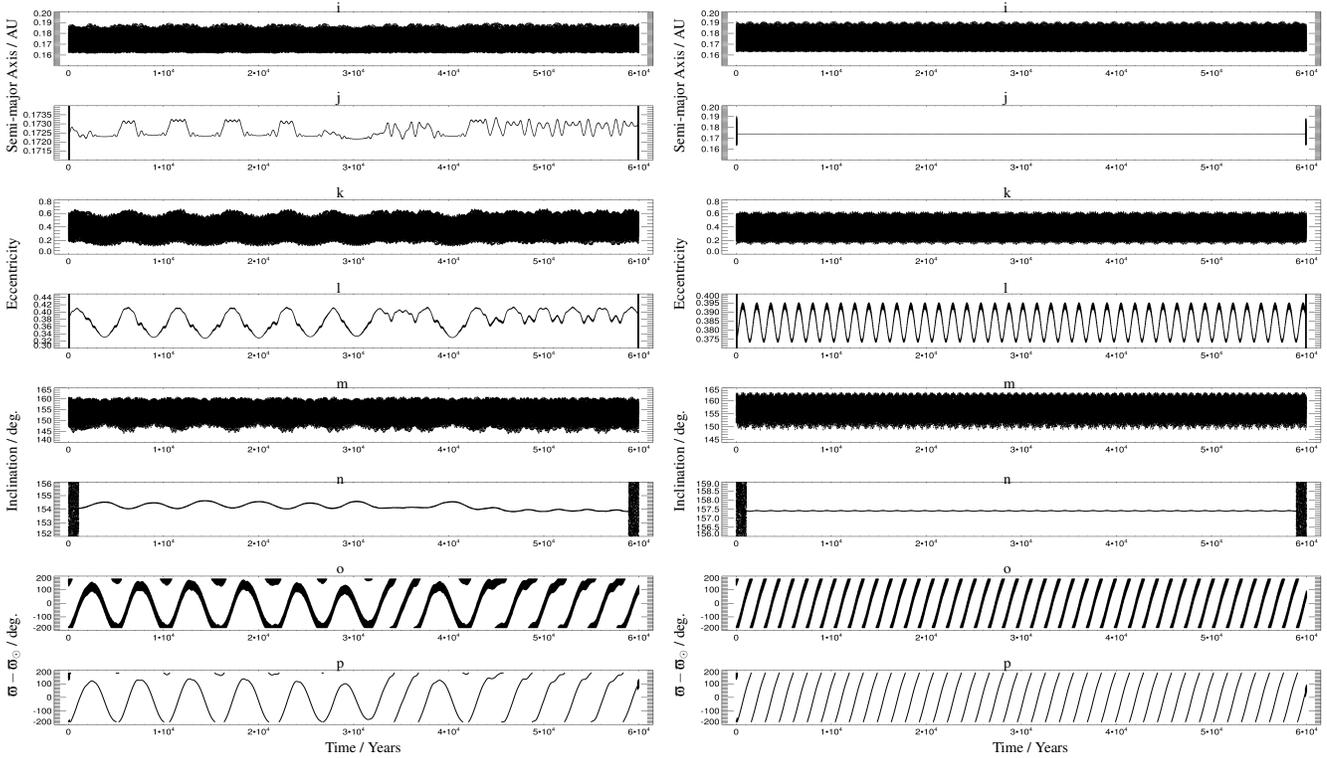

%\psdraft
%\centering
% Upper figure source:
%/home/tobiash/simulations/jupiter/SingleOrbitIntegrationsForTobias/MeanMotionResonances/5:1MMR/IC-4
%Comments: Figure generated with IDL: FigureForPaper_ExamplesOfSmoothing.pro
%Comments: Figure labels has been added using xfig (PaperPlot_SmoothExample.fig) 
% Lower figure source:
%/home/tobiash/simulations/jupiter/SingleOrbitIntegrationsForTobias/MeanMotionResonances/5:1MMR/IC-9
%Comments: Figure generated with IDL: FigureForPaper_ExamplesOfSmoothing.pro
%Comments: Figure labels has been added using xfig (PaperPlot_SmoothExample.IC-9.fig) 
% Note on how these figures where generated:
%The original EPS file was loaded into XFIG and the labels were added. Then
%using the XFIG export command "combined PDF/Latex" two files *.pdf and 
%*.pdf_t are generated. Then the PDF file is loaded into GIMP and the 
%resolution with the option to increase it to 500 pixels/inch. After the image
%was loaded into GIMP the images was safed as a PNG image using the same size as
%the PDF file. Then the *.pdf_t was renamed to *.png_t and the filename in
%*.png_t needs to be changed as well to include the PNG image.
\centerline{
        \hbox{
              \resizebox{88mm}{!}{\input PaperPlot_SmoothExample.png_t}
              \resizebox{88mm}{!}{\input PaperPlot_SmoothExample_IC-9.png_t}
              }
}
\caption{Time evolution of orbital elements for initial condition IC-I (left
panel, chaotic) and initial condition IC-II (right panel, quasi-periodic). The
full data set (panel a, c, e and g) and corresponding smoothed data set (panel
b,d,f and  h) are shown in each panel. From top to bottom panel: Semi-major
axis (a+b).  Eccentricity (c+d). Inclination (e+f) and the angle 
$\varpi-\varpi_{\odot}$ (g+h). The smoothing window applied to the data set
in both panels are identical. Only the initial conditions has changed. Half of 
the window widths for a given element can be seen as a black bar at the 
beginning and end  of the integration time. Each initial condition has been
integrated for 60000  years corresponding to the integration length in the
MEGNO maps. The window width are $w/2\times\Delta T=2500/2\times
40~\textnormal{days} = 137$ years  for panels b,d,g and $w/2\times\Delta T =
20000/2\times\Delta T = 1095$ years for panel f. Note that $\varpi_{\odot} = 
\varpi_{J}-180^{\circ}$.}
\label{SmoothedExample}
\end{figure*}

$A_{i}=A(t_{i})$ with $t_{i} = t_{0} + i\Delta T$  (for $i = 0,1, \ldots, N)$
the smoothed (secular) sequence ${R_{i}}$ is given by
\begin{displaymath}
R_{i} = \left \{ 
\begin{array}{ll}
\frac{1}{w} \sum_{j=0}^{w-1} A_{i+j-(\frac{w-1}{2})}; & 
\textrm{ $i = \frac{w-1}{2}, \ldots, N-\frac{w+1}{2}$.}  \\
A_{i} & \textrm{otherwise.}
\end{array} \right.
\label{SMOOTHequation}
\end{displaymath}
\noindent
Here $N$ is the number of data points from the numerical integration and $w$
is the (running) window width over within which the original data set is 
averaged on. In units of time the window width is simply $w\Delta T$. The
window width represents a free parameter and basically 
controls the supression of dynamical features seen in the data set: a too short
window will have little smoothing effect and thus the fast frequencies are
retained; a too large window width will suppress long-period features that
appear on secular time scales. In practice some experimentation is needed in
order to determine a satisfying window width. In this study we have performed
an extensive survey of various  window widths. In each plot showing smoothed
orbital elements we have chosen  the most appropriate window width in order to
highlight the secular changes of  a particular orbit.

Some care has to be taken when smoothing angular variables. As already mentioned
by \cite{SahaTremaine1993} filtering (or smoothing) an angular variable is
problematic as angles may change discontinuously over the time sequence (i.e
from $-\pi$ to $\pi$). Applying a running window smoothing  procedure directly
to an angular variable introduces spurious results around  discontinuities. To
circumvent this problem,  we transform a given angular variable to a continuous
signal. Let $\theta$ be an angular variable changing discontinuously at times 
during the numerical integration. Then we transform to the following continuous 
variables
\begin{eqnarray}
p &=& A \cos(\theta) \\
q &=& -A \sin(\theta),
\end{eqnarray}
where $A$ is some suitable constant. Here we chose $A=1$. The 
{\sc IDL} {\sc SMOOTH} smoothing procedure is then applied to each of
the quantities $p,q$ and we obtain  $\bar{p},\bar{q}$ after which we obtain
the smoothed angle $\bar{\theta}$ from $\bar{\theta} =
\arctan(\bar{q}/\bar{p})$.

An important note about the {\sc SMOOTH} routine is the following. The
averaging behaviour at the  beginning and end of the original time sequence
depends on the optional \texttt{edge\_truncate} keyword passed to the
{\sc SMOOTH} function. If this  keyword is enabled the smoothing procedure
might introduce false/misleading averages at  the beginning and end of the
smoothed signal. Details can be found in the {\sc IDL} documentation. In
addition if this keyword is disabled then it is  important to note that
$R_{i}=A_{i}$ for the first data  points up to (but not including) $(w-1)/2$
and $R_{i}=A_{i}$ from $N-(w+1)/2$  (but not including) to $N$.

We demonstrate the effect of successively applying time-averaging smoothing 
windows to the time evolution of the semi-major axis in 
Fig.~\ref{SmoothingSemiMajorAxis}. Fig.~\ref{SmoothingSemiMajorAxis}a shows the
whole signal over 60000 yrs with a sampling frequency of 40 days. At this stage
a secular period of about 2400 years is clearly present in the frequency
spectrum of the signal. Fig.~\ref{SmoothingSemiMajorAxis}b shows the first 200
years of the full data set (thin line) which is dominated by the orbital 
frequency of the satellite and an approximately 12 year period (Jupiter's 
orbital period). When applying two successive smoothing windows to the raw 
data with window width 3 yrs and 12 yrs we obtain the average signal 
overplotted as a thick line in Fig.~\ref{SmoothingSemiMajorAxis}b. 
Fig.~\ref{SmoothingSemiMajorAxis}c shows a zoom of the smoothed signal over 
the whole integration time (note the difference in range of the semi-major 
axis). Fig.~\ref{SmoothingSemiMajorAxis}d (thin line) shows the first 3500 yrs
of the previous signal. This time we detect a 34 year periodicity in the
semi-major axis. Applying a smoothing window removes the 34 yr period (thick
line in Fig.~\ref{SmoothingSemiMajorAxis}d) and the smoothed signal over the
60000 yrs is shown in Fig.~\ref{SmoothingSemiMajorAxis}e. Furthermore a 140 yr
period signal is present in the semi-major axis as shown in 
Fig.~\ref{SmoothingSemiMajorAxis}. Applying a fourth smoothing window to the
\begin{figure*}
%Figure info: 
%Located on: arpc2
%Path: /home/tobiash/simulations/jupiter/megno/LetterSimulations
%Original figure name: ai-space.2-figures.megno.eps
%Gnuplot script used: plot.ai-space.2-figures.megno.gp
%Manipulation with GIMP: *.eps loaded into GIMP (500 pix/in) and saved as PNG.
\includegraphics[width=177mm,height=78mm]{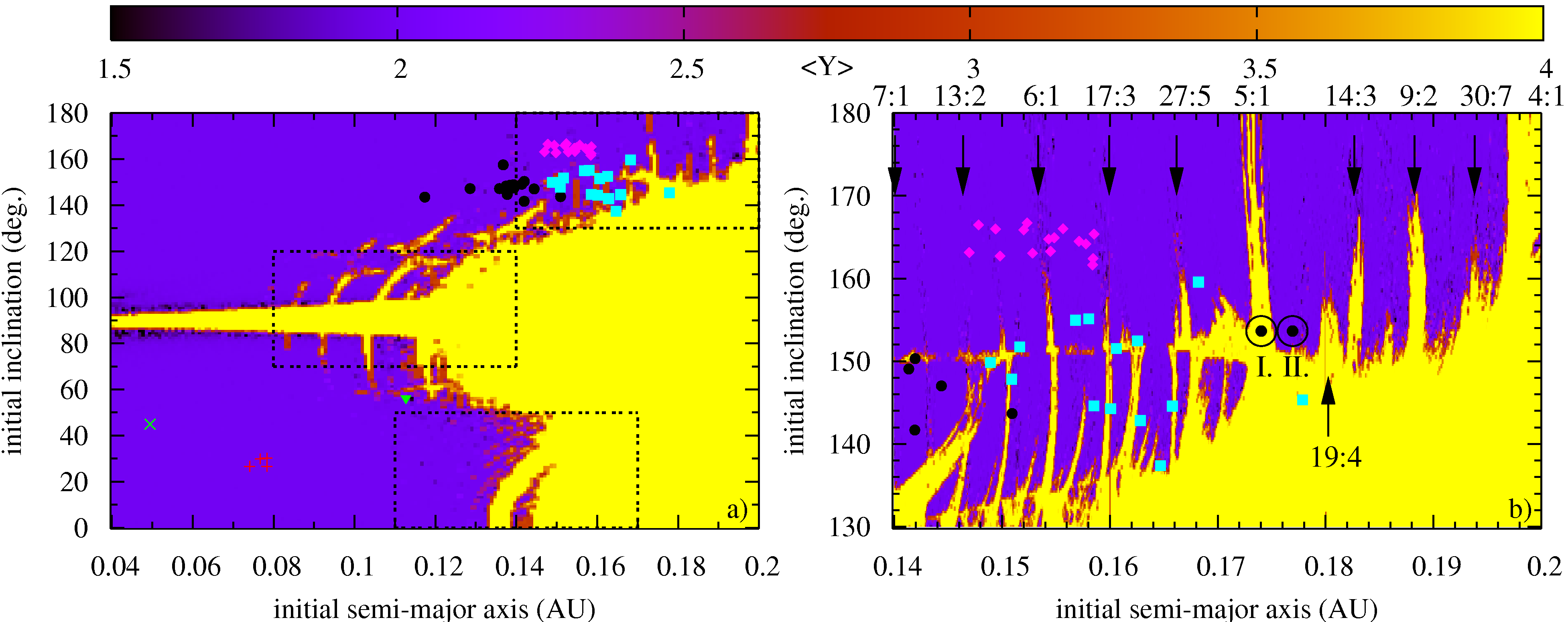}
\caption{MEGNO maps in $(a,I)$-space of test satellites with 
$e_{0}=0.20, \omega_{0}=0^{\circ}, \Omega_{0}=0^{\circ}, M_{0}=90^{\circ}$. 
The MEGNO indicator $\langle Y \rangle$ is color coded on a 
linear scale from 1.5 to 4
with $\langle Y \rangle = 2$ indicating quasi-periodicity and $\langle Y 
\rangle = 4$ denoting chaotic initial conditions. See the electronic version
for colors. In both panels the symbols denote locations of irregular
satellites. Themisto is at  0.05 AU ($0.14~\Rhill, 105~\Rjup$ and indicated by 
a $\times$ symbol). Himalia family = (+)-symbols and Carpo =
(triangle) at 0.112 AU $(0.32~\Rhill, 235~\Rjup)$. We plot family
members for which  $\vert e(t) - 0.20\vert < 0.05$. Ananke (black $\bullet$
symbols), Carme (pink rhomb symbols) and  Pasiphae family (magenta 
square symbols). \emph{Left panel a}:  Three regions showing
interesting dynamical features are shown by  rectangles. \emph{Right panel b}:
High resolution zoom of the upper right  region shown in panel a. Arrows
indicate the locations of retrograde mean-motion  resonances. Nominal locations
in are given in Table  ~\ref{TableOfMMRNominalLocation}. Also shown are the
locations (IC-I and  IC-II) of two test orbits. As already detected by
\citet{yokoyama2003} note  the small stable region at
$(a_{0},I_{0})=(0.142~\textnormal{AU},10^{\circ})$.}
\label{megnomaps}
\end{figure*}
original raw signal now also removes this period and we end up with the 
secular signal (2400 yr period) shown in Fig.~\ref{SmoothingSemiMajorAxis}f.

In Fig.~\ref{SmoothedExample} we show examples of the effect of the smoothing
procedure when applied to the numerical solution of two different initial 
conditions located close to each other in $(a,I)$-space. In both panels we
show the semi-major axis, eccentricity, inclination and the angle
$\varpi-\varpi_{\odot}$ for each initial condition. The left (right) panel is
the result of integrating initial condition IC-I (IC-II) as indicated in 
Fig.~\ref{megnomaps}. Each orbit were integrated for 60000 years, the 
integration length of each grid point in the MEGNO maps. The black bars (or 
sometimes scattered data points) shows half the window width (see figure 
caption for more details) and has been
experimentally  determined based on qualitative judgment with the goal to
enhance the  underlying dynamical changes in a given osculating element. As
already pointed out in \cite{SahaTremaine1993} we again observe that most of
the variation in the orbital elements are found in the fast frequencies. A
visual inspection of the time evolution of initial condition IC-I confirms
its chaotic nature as  correctly identified by MEGNO. For comparison a
time-running window with  identical window width has been applied to initial
condition IC-II. Its  quasi-periodic nature is clearly visible over the
considered time span. From experimentation with the window width we observed
that increasing the width averages out more and more quasi-periodic
oscillations. No chaotic behaviour has been observed when increasing the
window width. Following \cite[p.301]{MorbidelliNesvorny1999} by applying a
time-running  window smoothing procedure to a given osculating element we
obtain the corresponding \emph{proper element}. If the orbit evolves
quasi-periodically (regular motion with a limited number of frequencies) in time
then the  corresponding proper element is a constant of motion. On the other
hand if the corresponding proper element is chaotic then it exhibits a random
walk in  time. The chaotic nature of integrating initial condition IC-I
particularly  manifests itself in the time evolution of
$\varpi-\varpi_{\odot}$ as shown in Fig.~\ref{SmoothedExample}h (left panel).
This angle repeatedsly changes from libration to circulation; this is 
characteristic of chaotic behaviour (i.e perturbed pendulum model). For 
comparison this angle circulates for initial condition IC-II (quasi-periodic) 
over the entire integration time span without any change in oscillation mode.

\section{Results and discussion}

\subsection{MEGNO, $a_{max}$ and $e_{max}$ maps}
\label{stabilitymapsection}
Our results on computing the MEGNO indicator over a large grid in 
$(a_{0},I_{0})$-space is shown in Fig.~\ref{megnomaps}. The figure shows the
grid $(a_{0},I_{0}) \in ([0.04, 
0.20]~\textnormal{AU},[0.0^{\circ},180^{\circ}]$ region. The observed 
population of prograde and retrograde satellites are shown by various symbols
(see figure caption for details). We obtained the osculating elements of the 
irregular satellites from the JPL Horizon Ephemeris system (see earlier section
on initial conditions). We chose to consider this part of phase-space in order
to compare with previous results as published by
\cite{carruba2002}\footnote{The authors uses different units for the semi-major
axis. \cite{carruba2002} measures distance in units of Jupiter's  Hill radius
$\Rhill$, and \cite{yokoyama2003} measures distance in units of  Jupiter's 
radii $\Rjup$.} and \cite{yokoyama2003}. These authors explore similar phase
space  regions although they consider a much smaller grid of initial
conditions. Reproducing previously published results  motivated us to apply
and compute the  MEGNO indicator over a much larger region occupied by the
outer irregular Jovian satellites. In our work at initial time the
eccentricity was set to $e_{0}=0.20$ and the remaining Kepler elements
$(\omega_{0},\Omega_{0})$ were set to zero with $M_{0}=90^{\circ}$. We 
calculated the MEGNO indicator for 35100 initial conditions 
$(N_{x},N_{y})=(195,180)$. The orbit of each initial condition were integrated
for $5 \times 10^3 \Pjup$ years ($\Pjup$ is the orbital period of Jupiter).
\begin{figure*}
%Figure information:
%Location: arpc2 (armagh desktop computer)
%Path:/home/tobiash/simulations/jupiter/megno/LetterSimulations
%Gnuplotscript:plot.ai-space.2-figures.amax.WithZoomOfMMR.gp
%Original filename: ai-space.2-figures.amax.WithZoomOfMMR.eps
%Further notes: No manipulation with XFIG or Boundingbox.
%Further notes: GIMP was used to convert from *.eps to PNG.
\includegraphics[width=177mm,height=78mm]{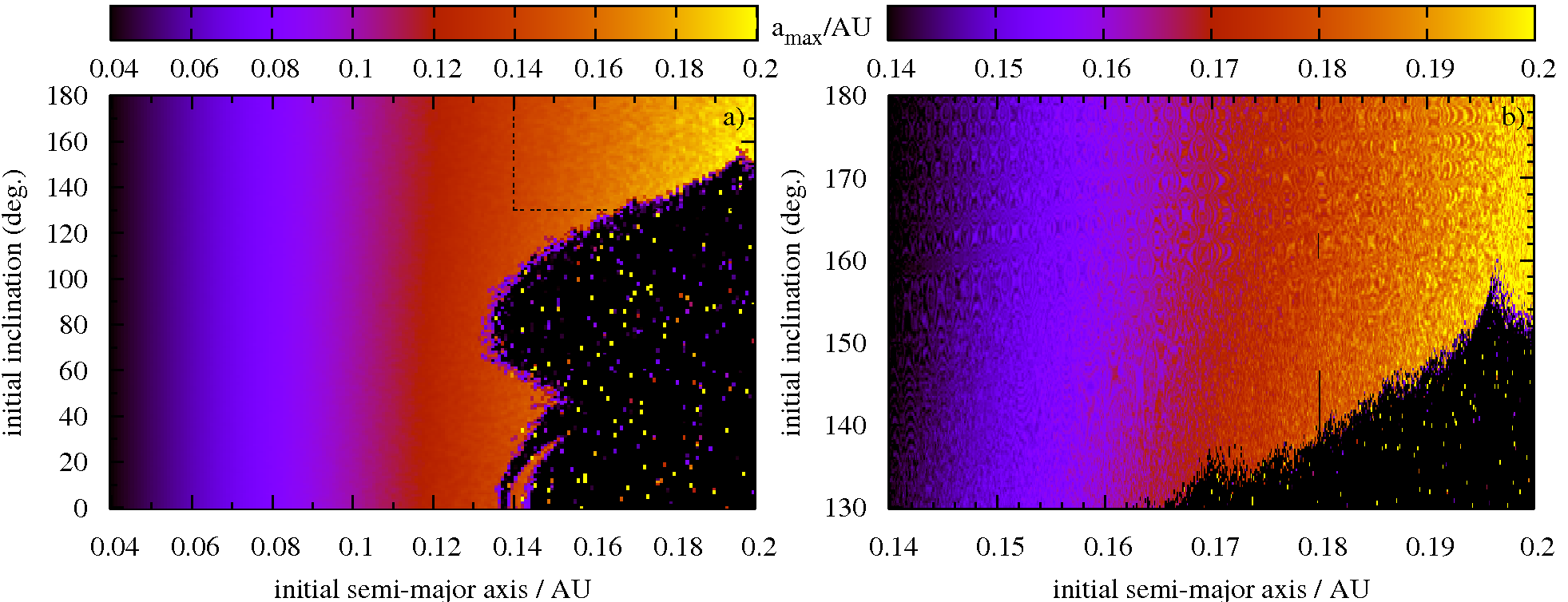}
\caption{Maximum semi-major axis ($a_{max}$) of a given initial condition in 
$(a,I)$-space for a irregular test satellite. \emph{Left panel} a: Same range 
in semi-major axis and eccentricity as for the MEGNO map with the maximum
semi-major axis color coded in the range $[0.04,0.20]$ AU $([0.11,0.56]~\Rhill, 
[84,419]~\Rjup)$. \emph{Right panel} b: Zoom of the rectangular area as 
indicated in the left a-panel figure.}
\label{amaxmaps}
\end{figure*}

\begin{table}
\begin{center}
\begin{tabular}{c c c}
\hline
\textnormal{MMR} ($n:n_{J}$) & $\textnormal{Nominal location (AU)}$ & 
\textnormal{Order} \\
\hline
\hline
 4:1 & 0.203250 & 3\\
\hline
30:7 & 0.194113 & 23\\
\hline
 9:2 & 0.187901 & 7\\
\hline
14:3 & 0.183400 & 11\\
\hline
19:4 & 0.181249 & 15\\
\hline
 5:1 & 0.175156 & 4\\
\hline
21:4 & 0.169550 & 17\\
\hline
16:3 & 0.167779 & 13\\
\hline
27:5 & 0.166396 & 22\\
\hline
11:2 & 0.164372 & 9\\
\hline
17:3 & 0.161133 & 14\\
\hline
23:4 & 0.159573 & 19\\
\hline
 6:1 & 0.155109 & 5\\
\hline
19:3 & 0.149618 & 16\\
\hline
13:2 & 0.147049 & 11\\
\hline
 7:1 & 0.139961 & 6\\
\hline
\hline
\end{tabular}
\caption{Nominal locations of retrograde MMRs as computed from 
Eq.~(\ref{MMRFormula}). The masses of Jupiter and the Sun and the semi-major 
axis of Jupiter are taken from Murray and Dermott (Table A.1 and A.2).}
\label{TableOfMMRNominalLocation}
\end{center}
\end{table}
To save computing time we stopped a given  integration as soon as $\langle
Y\rangle \ge  10.0$. The MEGNO indicator is then color-coded with $\langle
Y\rangle = 2.0$ indicating quasi-periodic motion and we only plot $\langle
Y\rangle \le 4$ to enhance the contrast of dynamical  features in the
transition region were the dynamics change from quasi-periodic to chaotic.
Figure ~\ref{megnomaps}a shows several interesting regions of chaotic  and
quasi-periodic nature with unprecedented detail. We decided to compute  high
resolution  maps to study this interesting region. A zoom plot is shown in 
Fig.~\ref{megnomaps}b and corresponds to the area within the box in the upper 
right corner of Fig.~\ref{megnomaps}a. The labels IC-I and IC-II correspond to
initial conditions for two hypothetical (retrograde) irregular satellites.

\begin{figure*}
%\psdraft
\centering
%Figure information:
%Location: arpc2 (armagh desktop computer)
%Path:/home/tobiash/simulations/jupiter/megno/LetterSimulations
%Gnuplotscript: plot.ai-space.2-figures.emax.WithZoomOfMMR.gp
%Original filename: ai-space.2-figures.emax.WithZoomOfMMR.eps
%Further notes: No manipulation with XFIG or Boundingbox.
%Further notes: GIMP was used to convert from EPS to PNG (500 pix/in).
\includegraphics[width=177mm,height=78mm]{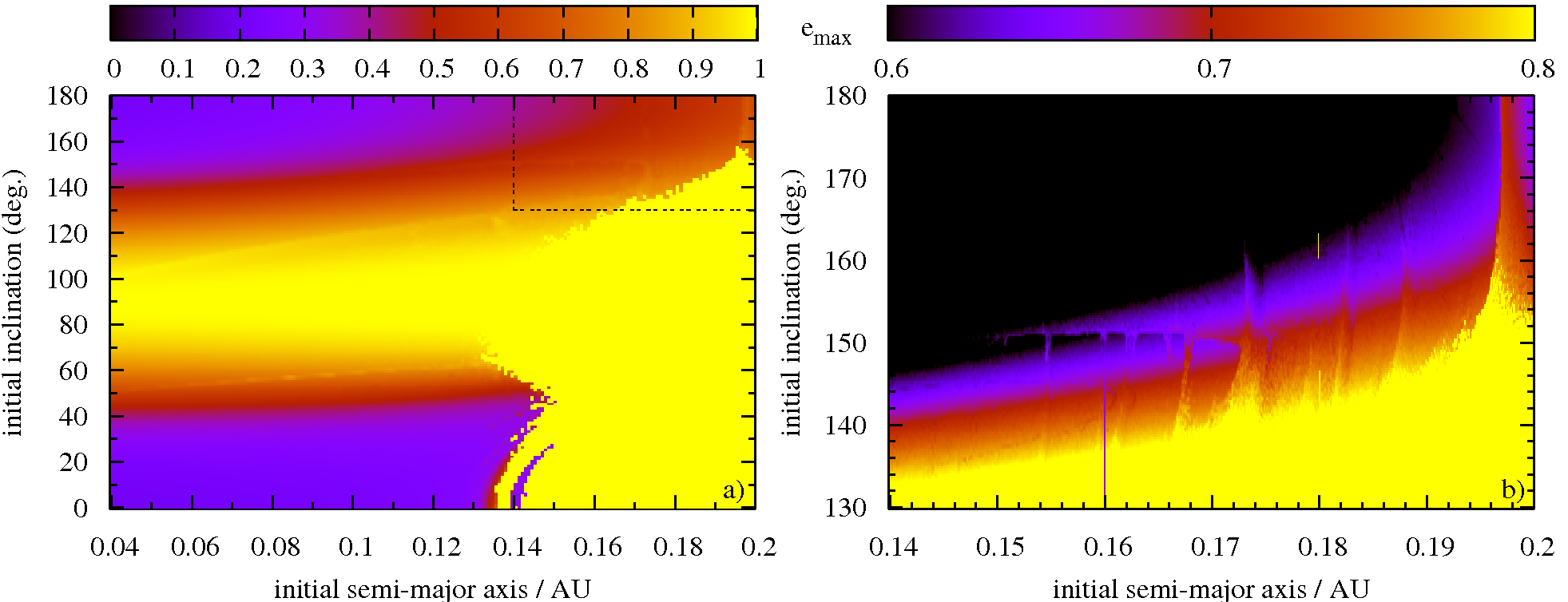}
\caption{Same as Fig.~\ref{amaxmaps} but using $e_{max}$ for the colorcoding.
Note the horizontal feature of enhanced eccentricity for both pro- and
retrograde satellites. This feature was also reported to exists from a 1 Myr
integration \citep[fig.6]{yokoyama2003}.}
\label{emaxmaps}
\end{figure*}
 
In parallel with computing $\langle Y \rangle$ to detect chaos we also  recorded
the maximum values of the test satellites semi-major axis and  eccentricity
every orbital period of the Sun. Fig.~\ref{amaxmaps} and  \ref{emaxmaps} show
the corresponding maps. It is now apparent that the global chaotic region
results in either escape or collisions after only $5 \times 10^3 \Pjup$ years.
From this result we can conclude that the large chaotic region detected in the
MEGNO map is strongly correlated with either escape from the Jovian Hill sphere
or collisions with Jupiter itself. It is  also interesting to note that over a
large range the orbit size (semi-major  axis) remains close to its initial
value. Furthermore we also  note that the locations of mean motion resonances
are not visible in the zoom plot (Fig.~\ref{amaxmaps}b) as otherwise indicated
by MEGNO. This  indicates that the semi-major axis is unaffected by the
presence of mean  motion resonances and MEGNO is efficient in detecting mean
motion resonances.  A different picture is obtained when studying the maximum
eccentricity  attained during a numerical integration of a test satellite. 
Fig.~\ref{emaxmaps} shows a clear dependence of $e_{max}$ with inclination.  In
a later section we will study and discuss the time evolution of these  initial 
conditions to unravel their chaotic and quasi-periodic nature in detail
presenting a proof-of-concept of detecting chaotic initial conditions. In this
work we will discuss the location of mean motion resonances and postpone the 
study and discussion of other features (boxes located at polar and prograde 
orbits) in a future work.

\subsection{Fine structure of mean-motion resonances}
Fig.~\ref{megnomaps}b shows details of the retrograde region with 
$a\in[0.14;0.20]$ AU $([0.39,0.56]~\Rhill, 
[293,419]~\Rjup)$ and $I \in [130^{\circ};180^{\circ}]$. Several vertical
structures are observed corresponding to the location of chaotic 
mean-motion resonances. If $n$ and $n_{J}$ denotes the mean-motion of the 
satellite and Jupiter respectively, then the semi-major axis of a satellite  
in a $n:n_{J}$ mean-motion resonance with Jupiter is given by
\begin{equation}
a_{s} = a_{J} \Bigg (\frac{n}{n_{J}}\Bigg)^{-2/3} 
\Bigg ( \frac{m_{J}+\Msun}{m_{J}}\Bigg)^{-1/3},
\label{MMRFormula}
\end{equation}
\noindent
where $a_{J},m_{J}$ denote Jupiter's semi-major axis and mass, respectively. 
$\Msun$ is the mass of the Sun. In the figure we show the location of several 
mean-motion resonances by arrows. See Table \ref{TableOfMMRNominalLocation} 
for their nominal locations. Several high-order mean-motion resonances are 
detected. 

When compared to the current population of observed irregular satellites it  is
interesting to note the large scatter in $(a,I)$ elements of the Pasiphae 
group. Fig.~\ref{megnomaps}b strongly implies that several  members of this
group are strongly affected by the dynamics within high-order  mean-motion
resonances. The  dynamical consequences of mean-motion resonances on the
orbits of irregular  satellites were already reported by 
\cite{SahaTremaine1993} discussing the $n-6n_{J} \sim 0$ resonance of  Sinope
and possibly S/2001 J11 \citep{nesvorny2003}. Showing a more compact
distribution of the osculating elements members of the Carme family are on
less inclined orbits  $I \approx 165^{\circ}$. It appears that the dynamical
effects of the  mean-motion resonances occurs at a smaller magnitude for this
group possibly  having the effect of a smaller dispersion in orbital elements.
In addition  the Ananke group shows also a small scatter in their orbital
elements  possibly due to the close proximity of the 7:1 mean-motion
resonance. The coincidence between the scatter of orbital elements of
retrograde  satellites and the presence of mean-motion resonances is probably
not by chance. We stress that this work does not address the dynamical
significance and effects of mean-motion resonances. Our results raises the
question of the dynamical  effects of mean-motion resonances on the orbital
elements on a compact group  of satellites. We will address this question in a
future work and study the  distribution of orbital elements of an initial
compact group by gravitational  scattering in mean-motion resonances.

\subsection{Chaotic regions and Kozai mechanism}

\begin{figure}
%Figure location on arpc2:
%IC-4:/home/tobiash/simulations/jupiter/SingleOrbitIntegrationsForTobias/MeanMotionResonances/5:1MMR/Orbit5Chaotic
%IDL Plot: FigureForPaper_5TO1MMR.pro
%NOTE: Labels were added using XFIG (special flag, export EPS/Latex). The
%*.pstex file was then loaded into GIMP and converted into PNG.
%\resizebox{86mm}{!}{\input test.pdf_t}
\resizebox{86mm}{!}{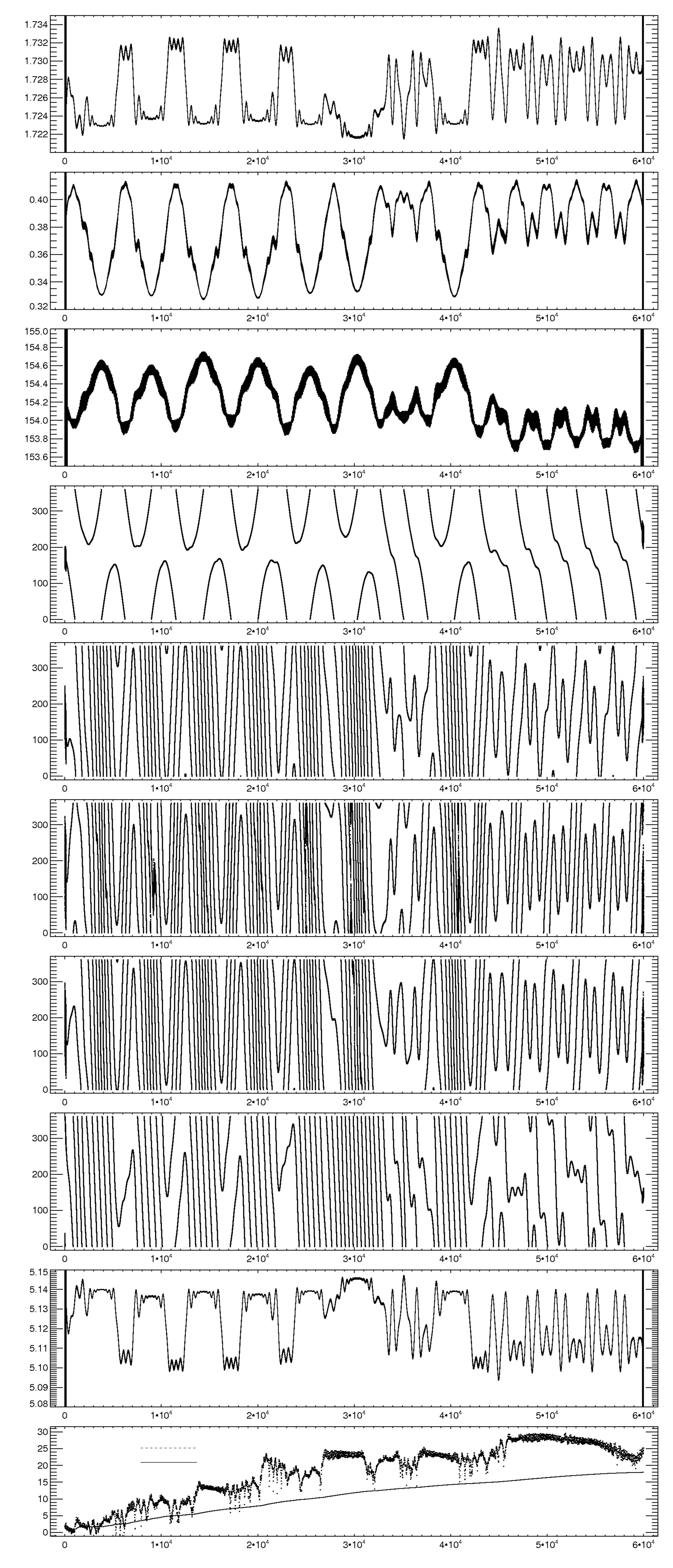}
\caption{Secular time evolution of osculating elements from integrating IC-I. 
Elements $a,e$ were smoothed using a window width of $w\times\Delta T = 2500 
\times 40~\textnormal{days} = 274$ years. For the inclination $I$ we used 
$w=5000$. The black vertical bars shows half the window width. The elements 
$\varpi, \lambda$ denotes the (retrograde) longitude of pericenter and mean 
logitude of the satellite, respectively. $\rm P$ measures the orbital period of
the satellite and $\Psun$ is the orbital period of the Sun in the jovicentric
system.
$\phi_{1}=5\lambda_{\odot}-\lambda-4\varpi_{\odot}$.
$\phi_{2}=5\lambda_{\odot}-\lambda-2\varpi_{\odot}-2\varpi$.
$\phi_{3}=5\lambda_{\odot}-\lambda-\varpi_{\odot}-\varpi-2\Omega_{\odot}$.
$\phi_{4}=5\lambda_{\odot}-\lambda+2\varpi-6\Omega_{\odot}$.}
\label{Orbit5Chaotic}
\end{figure}

From the MEGNO map (Fig.~\ref{megnomaps}a), we observe a clear distinction
between quasi-periodic and chaotic phase space regions. The general chaotic
region for prograde satellites starts  from $a=0.14~\textnormal{AU}~
(0.39~\Rhill,293~\Rjup)$ and  outwards. Stable quasi-periodic orbits for the
retrograde satellites are found for orbits with semi-major  axis up to 0.195 AU
$(0.55~\Rhill, 409~\Rjup)$. It is interesting to note that the chaotic
phase-space of prograde satellites is larger when  compared to the retrograde
satellites.  This asymmetry in $(a,I)$ space was already discussed by
\cite{nesvorny2003} pointing out the difference between prograde and retrograde
stability limits.  At larger semi-major axis the retrograde satellites are on
dynamically more  stable orbits when compared to the prograde irregular
satellites. In this work the calculated MEGNO map confirms this stability
asymmetry which is mainly explained by the existence of the evection resonance
for prograde satellites  \citep[and references
therein]{nesvorny2003,yokoyama2008}. In geometrical terms this resonance locks the satellites apocenter towards the 
direction of the Sun and maintains this alignment for a period of time. In this
orbital configuration solar perturbations accumulate and the satellite
eventually escapes from Jupiter  when $\varpi-\lambda_{\odot} \sim 0$. An 
interesting feature shown in the map is the chaotic `horizontal cone' for polar
orbits at around $I \sim 90^{\circ}$ extending from
$a=0.04~\textnormal{AU}~(0.11~\Rhill, 84~\Rjup)$ to 
$a=0.12~\textnormal{AU}~(0.34~\Rhill, 251~\Rjup)$. This region was already
studied intensively by  \cite{carruba2002,nesvorny2003} showing that test
satellites with  inclinations in the range $70^{\circ} < I < 110^{\circ}$ are
short-lived  orbits with survival times less than $10^{7}$ years. Analytical
perturbation theory \citep{carruba2002} showed that the accumulation of
secular solar perturbations is the driving force causing the excitation of

\begin{figure}
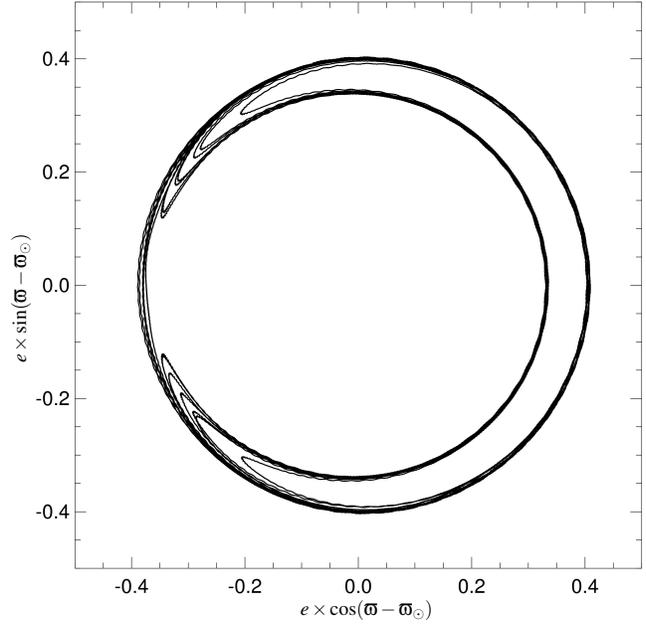

%Figure location on arpc2:
%IC-4:/home/tobiash/simulations/jupiter/SingleOrbitIntegrationsForTobias/MeanMotionResonances/5:1MMR/IC-4
%IDL Plot: PlotSMOOTH_POLARPLOT.pro
%NOTE:XFIG were used to add the axis labels then exported using EPS/Latex both
%parts. Then the *.pstex file was converted to PNG and the *.pstex_t renamed to
%*.png_t and the include filename changed.
\resizebox{84mm}{!}{\input SmoothingPolar_ecc.png_t}
\caption{Polar representation of the temporary circulation and librations 
of $\varpi-\varpi_{\odot}$ about $0^{\circ}$ 
($\varpi-\varpi_{J}=180^{\circ}$) in the 5:1 mean-motion resonance of IC-I. 
The outer radius corresponds to $e \approx 0.33$ and the inner radius 
corresponds to $e\approx 0.41$. The alternation between libration and 
circulation is a clear sign of chaotic dynamics.}
\label{5TO1MMR-PolarPlot}
\end{figure}
\noindent
orbit eccentricity to large values. This mechanism is known as the Kozai
resonance. Depending on the initial value of $\varpi$ circular orbits can reach
high eccentricities and collisions with  the inner regular satellites (or with
Jupiter itself) are expected to occur. Thus the Kozai resonance  provides an
efficient dynamical  mechanism to remove an initial population of near-polar
irregular satellites. In near-polar orbits when studying Fig.~\ref{emaxmaps} we
 notice  that the Kozai regime is well identified with large eccentricities
attained in  the range $40^{\circ}<I<140^{\circ}$. However, it needs to be
mentioned that the eccentricity excitations might  strongly depend on the
initial $\varpi$ \citep{carruba2002}.

\subsection{Chaoticity versus quasi-periodicity}

Previously we have reported on the presence of chaotic dynamics as detected
from calculating $Y,\langle Y\rangle$ on a grid of initial conditions in 
$(a,I)$-space. In the following we will study in detail two initial 
conditions that were detected to be either quasi-periodic or chaotic. 
Fig.~\ref{megnomaps} marks the two initial conditions by dots (surrounded by 
circles) with IC-I and IC-II. Both initial conditions have the same orbit
inclinations with different semi-major axis. We refer to Table
\ref{variousICs} for details on numerical values in the initial conditions. We
have then numerically  integrated both orbits using the Radau integrator with
initial step size of 0.01 days and tolerance parameter of $10^-13$.
Calculations were done in double precision enabling the `high' output
precision in {\sc MERCURY}. Initial conditions for the Sun were obtained
from the JPL Horizon Ephemeris.  We sampled osculating elements every 40 days.
To maintain consistency we  integrated the orbits over $5 \times 10^3~\Pjup$
years. For both orbits the  maximum relative energy error at the end of 
integration were smaller than  $10^{-13}$.

\begin{figure}
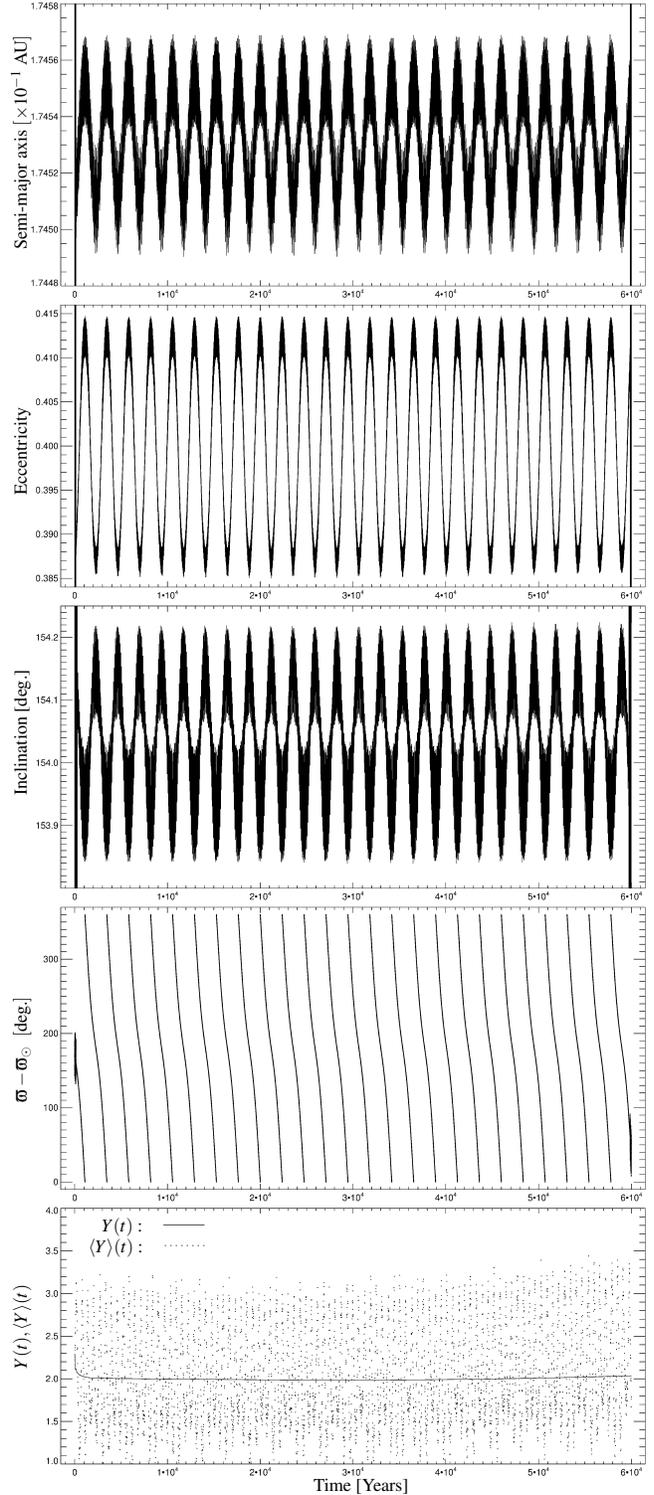

%Figure location on arpc2:
%/home/tobiash/simulations/jupiter/SingleOrbitIntegrationsForTobias/MeanMotionResonances/5:1MMR/IC-4
%IDL Plot: PlotSMOOTH_POLARPLOT.pro
%NOTE:XFIG were used to add the axis labels then exported using EPS/Latex both
%parts. Then the *.pstex file was converted to PNG and the *.pstex_t renamed to
%*.png_t and the include filename changed.
\resizebox{84mm}{!}{\input Orbit5Quasiperiodic_NewBoundingBox.png_t}
%\scalebox{0.58}{\input{./Orbit5Quasiperiodic.NewBoundingBox.ReducedQualityByGIMP.pstex_t}}
\caption{Time evolution of a test satellite with initial condition IC-II. The
same window width were used as in Fig.~\ref{Orbit5Chaotic}. Note: The range in
semi-major axis, eccentricity and inclination is smaller than in
Fig.~\ref{Orbit5Chaotic}. This time the angle $\varpi-\varpi_{\odot}$ is
circulating as opposed to the chaotic case.}
\label{Orbit5Quasiperiodic}
\end{figure}

\begin{table}
\begin{center}
\begin{tabular}{l c c c c}
\hline
Initial conditions & $a_{0}~(\textnormal{AU})$ & $e_{0}$ & 
$I_{0}~(^{\circ})$ & Region\\
\hline
\hline
IC-I   & 0.173915  & 0.20 & 153.968 & CH \\
IC-II  & 0.176179  & 0.20 & 153.968 & QP \\
\hline
\hline
\end{tabular}
\caption{Initial conditions for the two test satellites IC-I and IC-II in and
close to the 5:1 mean-motion resonance as shown in Fig.~\ref{megnomaps}. The 
remaining Kepler elements were set to $\omega_{0}=0^{\circ}$, 
$\Omega_{0}=0^{\circ}$ and $M_{0}=90^{\circ}$. CH and QP denotes 
initial conditions in the chaotic region and quasi-periodic region, 
respectively.}
\label{variousICs}
\end{center}
\end{table}

As suggested from MEGNO initial condition IC-I exhibits chaotic behaviour. We
present the results of our single orbit calculation in 
Fig.~\ref{Orbit5Chaotic} and Fig.~\ref{5TO1MMR-PolarPlot}. We have obtained
those figures by successively applying the running window time average
smoothing technique as outlined previously. In addition to the orbital
elements  $(a,e,I)$ we also plot the time variation of 
$\varpi-\varpi_{\odot}$ along with four critical angles 
$\phi_{1},\phi_{2},\phi_{3},\phi_{4}$ (see details in the figure caption). In
the bottom panel of Fig.~\ref{Orbit5Chaotic} we also show the time evolution
of $Y,\langle Y\rangle$ as a function of time. We obtained this plot from a
GBS integration. Quantitatively at time 60000 yrs $\langle Y\rangle \approx 
18$ and a visual inspection shows a clear trend of divergence of  $\langle
Y\rangle$ over time. The presence of chaos is qualitatively best  seen in the
time evolution of the angle $\varpi-\varpi_{\odot}$. In Fig.
~\ref{5TO1MMR-PolarPlot} we give a polar representation of this angle. A
similar approach was also adopted by \cite{SahaTremaine1993}. In the 
beginning this angle librates around $\varpi-\varpi_{\odot}=0^{\circ}$. After 
approximately 30000 years this libration mode switches into circulation for  a
short time period and then returns to the libration mode. At the end the 
angle circulates. This qualitative change between different modes of motion is
characteristic of motion near a separatrix and hence chaotic motion is
concluded. In the polar representation times of temporary librations are shown
as `banana'-shape curves and circulations are indicated by full circles.

A more elongated banana corresponds to a larger libration amplitude of
$\varpi-\varpi_{\odot}$. From Fig.~\ref{Orbit5Chaotic} it is apparent that the
libration amplitude changes with time. The initial conditions are chosen with
the orbital inclination to be initially outside the Kozai resonance for which
$\omega$ of the satellite starts to librate around either $90^{\circ}$ or
$270^{\circ}$ \citep{carruba2002,nesvorny2003,yokoyama2003}. No large
eccentricity variations are expected. It is important to note the difference 
of the location of the libration centre found for the test satellite started
at IC-I when compared to the libration behaviour of this angle for several
major satellites. \cite{SahaTremaine1993,WhippleShelus1993} report that for 
the retrograde satellites Pasiphae and Sinope the angle $\varpi -\varpi_{J}$ 
librates about $180^{\circ}$. The difference from this work is the choice in
reference system. In a heliocentric system the angle  $\varpi -\varpi_{\odot}$
for IC-I would librate about $-180^{\circ}$ which is consistent with the
general trend as reported in \cite{SahaTremaine1993}.

\begin{figure*}
%Figure information:
%Location: arpc2
%Path:/home/tobiash/simulations/jupiter/megno/data/a-i-scans@e-constant
%Gnuplot script:PlotMegno.e=0.10.PolarPlot-Paper.NoAnnotations.gp
%Gnuplot script:PlotMegno.e=0.20.PolarPlot-Paper.NoAnnotations.gp
%Note: The eps file from gnuplot has been edited to crop the bounding box.
%Then the edited eps figure has been loaded into xfig.
%XFIG script:megno.JaiRun6.e=0.10.PolarPlot-Paper.NoAnnotations.fig.
%XFIG script:megno.JaiRun1.e=0.20.PolarPlot-Paper.NoAnnotations.fig.
%\scalebox{0.78}{\input{./megno.JaiRun6.e=0.10.PolarPlot-Paper.NoAnnotations.QualityReducedByGIMP.pstex_t}}
%\scalebox{0.78}{\input{./megno.JaiRun1.e=0.20.PolarPlot-Paper.NoAnnotations.QualityReducedByGIMP.pstex_t}}
%\scalebox{0.78}{\input{./megno.JaiRun6.e=0.10.PolarPlot-Paper.NoAnnotations.pstex_t}}
%\resizebox{88mm}{!}{\input megno_JaiRun6_e=0_10_PolarPlot-Paper_NoAnnotations.png_t}
%\resizebox{88mm}{!}{\input megno_JaiRun1_e=0_20_PolarPlot-Paper_NoAnnotations.png_t}
\resizebox{88mm}{!}{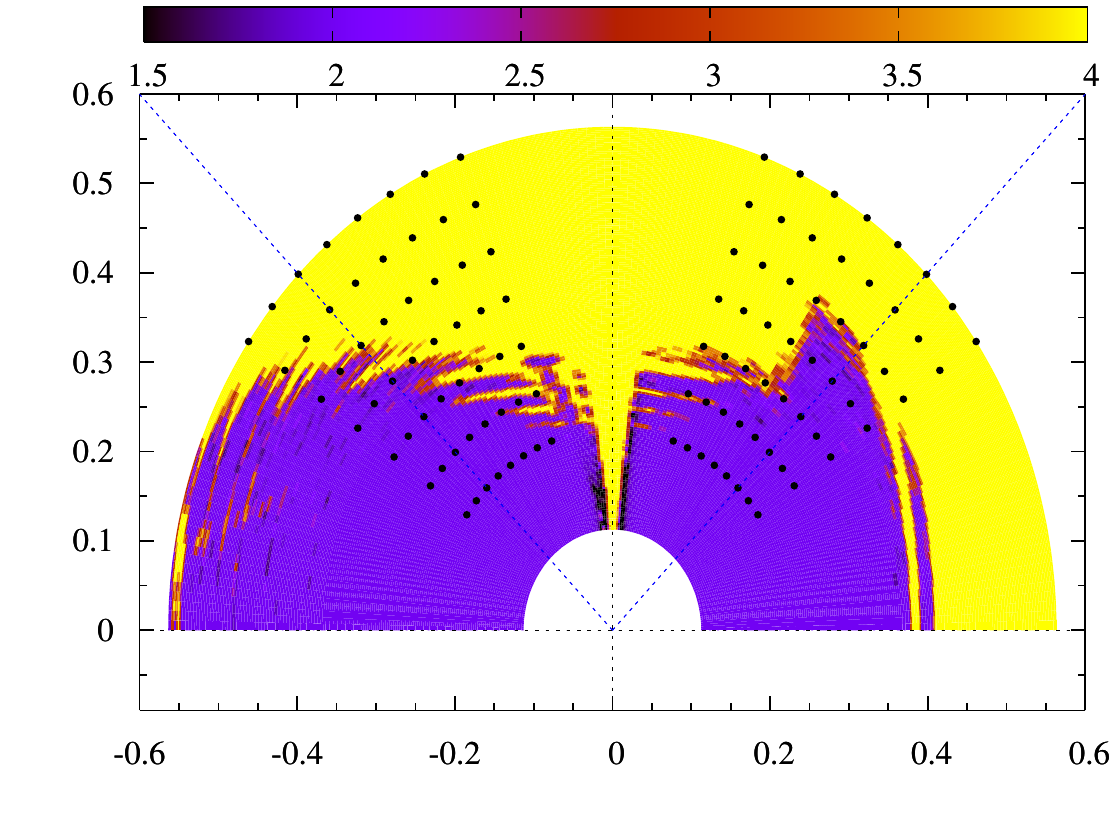}
\resizebox{88mm}{!}{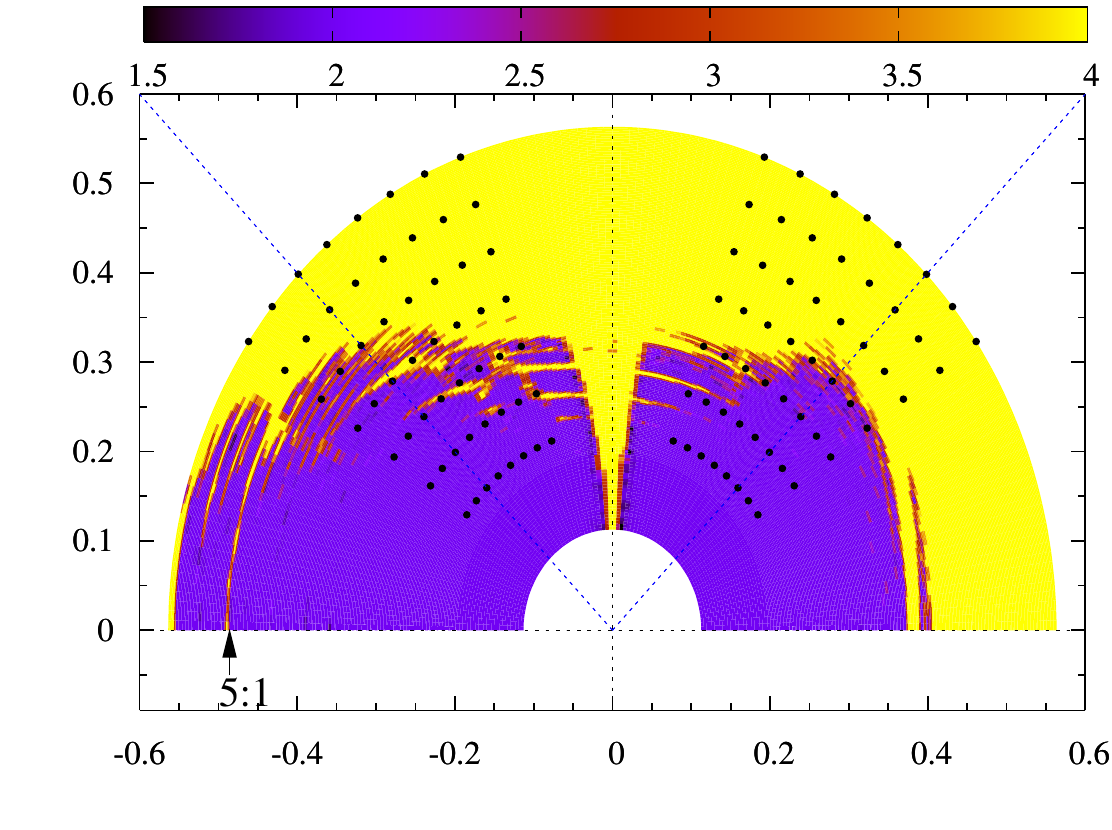}
%\scalebox{0.85}{\includegraphics{./megno.JaiRun1.e=0.20.PolarPlot-Paper.eps}}
\caption{Polar representation of the $(a,I)-$space of Jovian irregular
satellites for two different initial orbit eccentricities. \emph{Left panel}:
$e_{0}=0.10$. \emph{Right panel}: $e_{0}=0.20$. In both panels the MEGNO indicator 
$\langle Y \rangle$ is color coded from 1.5 to 4.0 (see the electronic version
for colors). The initial semi-major axis $(a_{0})$ is given in units of 
Jupiters Hill radius $\Rhill$, and $I_{0}$ denotes the initial orbit 
inclination. The diagonal lines represents orbit inclinations at $45^{\circ}$ 
(prograde satellites) and $135^{\circ}$ (retrograde satellites). The central
vertical line represents the polar orbit at $I_{0}=90^{\circ}$. For comparison 
black dots indicate the initial conditions studied in \citep{carruba2002}.}
\label{polarplots}
\end{figure*}

Since IC-I is near the 5:1 mean-motion resonance 
(compare with Table~\ref{TableOfMMRNominalLocation}) we have systematically
explored and looked for the existence of critical angles of the form
\cite{MurrayDermott2001,MorbidelliBook2002,yokoyama2003}
\begin{equation}
\phi = k_{1}\lambda_{\odot}+k_{2}\lambda+k_{3}\Omega_{\odot}+k_{4}\Omega+
k_{5}\varpi_{\odot}+k_{6}\varpi,
\end{equation}
\noindent
with $\sum k_{i} = 0$ and the sum of the coefficients of the nodes being even
as is required by the d'Alembert's rules. The strength of a given critical
angle depends on the power in eccentricity. From Fig.~\ref{Orbit5Chaotic} we have  $a_{max}$ then
$\phi_{1},\phi_{2},\phi_{3},\phi_{4}$ circulates prograde sense. 

Also whenever $I$ is minimum (maximum) then $e$ is maximum (minimum). Also
whenever $\phi_3$ librates around $\pi$ then $I$ is at a minimum and constant
(at time 35000 years). We also see that at the end of the time evolution of the
satellite the angles $\phi_{1},\phi_{2}$ and $\phi_{3}$  are switching between
libration around $\pi$ and prograde circulation. At some times we see a
temporary resonance lock in  $\phi_2$ in the 5:1 mean-motion resonance.

Turning our attention to the time evolution of the quasi-periodic orbit 
started at IC-II we report on the following results.
Fig.~\ref{Orbit5Quasiperiodic} shows the smoothed time evolution of the
osculating elements $(a,e,I)$ along with $\varpi-\varpi_{\odot}$. In the
bottom panel plot we again plot the time evolution of $Y,\langle Y \rangle$
over 60000 years. The smoothing follows the running window time average 
techniques as outlined previously. Contrary to the chaotic orbit the angle
$\varpi-\varpi_{\odot}$ now cirulates only and the secular time variation of 
the semi-major axis, eccentricity and inclination are characterised by
quasi-periodic oscillations. 

Based on the preceeding comparative study we conclude that MEGNO is a reliable
numerical tool for detecting chaotic dynamics on short time scales. We plan to 
use this tool in future work on the dynamics of irregular satellites for the
major planets in the Solar System. A particular interesting subject of study
would be the change of mass of Jupiter and the corresponding change in the
topology structure of phase-space at a given time during the growth phase of 
Jupiter.

\subsection{Comparing with previous work}

In the following we compared our MEGNO maps with numerical studies published 
previously in the literature. Fig.~\ref{polarplots} shows a polar
representation of Fig.~\ref{megnomaps}a for two different values of initial
eccentricities $e_{0}=0.10$ and $e_{0}=0.20$. In Fig.~\ref{polarplots}a,b the
final value of $\langle Y \rangle$ after 60000 yrs of integration time is
color coded with yellow indicating  chaotic dynamical time evolution and blue
indicating initial conditions exhibiting quasi-periodic dynamics. The
dynamical and collisional evolution of irregular satellites and their 
lifetimes has been studied by \cite{carruba2002} and \cite{nesvorny2003}. We 
have compared Fig.~\ref{polarplots} with the results of long-term integrations
 of hypothetical irregular satellites conducted by 
\citet[fig.~8, fig.~9]{carruba2002}. Similar studies can be found in
\cite[fig.~9]{nesvorny2003}. Initial conditions  of test satellites studied in
\cite{carruba2002} have been superimposed by black dots in both maps following
the exact array of initial conditions as presented in \cite{carruba2002}. The
initial semi-major axis ranges from 0.08 AU to 0.20 AU with spacing $\Delta a
= 0.02~\textnormal{AU}$. The initial inclination is $35^{\circ}$ to
$70^{\circ}$ for the prograde and $110^{\circ}$ to  $145^{\circ}$ for
retrograde satellites with spacing $\Delta i=5^{\circ}$. In  the following
discussion it is important to stress the difference in the dynamical models
used. In this work we only consider the  Sun-Jupiter-test particle system
without considering collisions with other bodies or ejections. 
\cite{carruba2002} includes the perturbative effects of the major planets and 
includes collision and ejection criteria within Jupiter's Hill sphere.

When comparing our results for prograde test satellites ($e_{0}=0.10$) with 
\cite[fig.~8]{carruba2002} all test particles with lifetimes less than 10 Myrs
$(a_{0} \ge  0.14~\textnormal{AU})$ are located in (or close to the onset of)
the chaotic region shown in Fig.~\ref{polarplots}a. Initial conditions started
in the quasi-periodic region have stable orbits over 1 Gyrs for
low-inclination $(I_{0} \le 50^{\circ},~a_{0} \le 0.12~\textnormal{AU})$ 
orbits to 10 Myrs for high-inclination orbits $(I_{0} \ge 65^{\circ},~a_{0} 
\ge 0.08~\textnormal{AU})$. It is important to note that although our MEGNO map
shown in Fig.~\ref{polarplots}a indicates quasi-periodic dynamics for the 
high-inclination orbits (for $a_{0}=0.08~\textnormal{AU}~\textnormal{to}~ 
0.12~\textnormal{AU}$) those initial conditions are short-lived due to the
presence of the Kozai cycle which opens up a route for those satellites to
reach into the region of the orbits of the Gallilean satellites. As
demonstrated by \cite{carruba2002} such high-inclination test satellites will
experience close encounters or  collisions with the massive regular satellites
of Jupiter. For the retrograde orbits (with $e_{0}=0.10$) a comparison allows 
the conclusion that all test satellites started in (or close to) the chaotic
region have lifetimes smaller than 1 Gyrs and all test satellites with initial
conditions in the quasi-periodic region have lifetimes over 1 Gyr with the
exception of high-inclination orbits exhibiting Kozai cycles in their orbital
eccentricities and inclinations. Similar conclusion are obtained when 
comparing the results shown in Fig.~\ref{polarplots}b $(e_{0}=0.20)$ with 
\cite[fig.~8]{carruba2002}. A final interesting point to mention is the 
qualitative difference in $(a,I)$-phase space topology of irregular satellites 
when varying the initial eccentricity. The prograde quasi-periodic region 
seen in Fig.~\ref{polarplots}a at $I_{0} \le 35^{\circ}$ with 
$a_{0}/\Rhill \times \cos(I_{0}) \approx 0.4$ has significantly 
decreased in Fig.~\ref{polarplots}b when $e_{0}=0.20$. Furthermore the 5:1 MMR
for retrograde orbits manifests itself more prominently in
Fig.~\ref{polarplots}b when compared to Fig.~\ref{polarplots}a. This suggests
that the structure of  $(a,I)$-phase space topology is strongly dependent on
initial eccentricity  and a survey of this region of parameter space is 
currently ongoing.

\section{Discussion and conclusions}

We have introduced, described and applied the MEGNO chaos indicator to the
dynamics of jovian irregular satellites. Our results are based on the elliptic
restricted three-body problem considering a test particle (irregular satellite)
in a jovian planetocentric reference system perturbed by the Sun on an elliptic
orbit. Initial conditions for the numerical integrations of the orbits have
been obtained  from the JPL Horizon Ephemeris Generator. Basic concepts and
properties of  MEGNO are reviewed and described as well as details on the
practical computation of $Y(t)$ and $\langle Y \rangle(t)$ based on the
variational equations. Numerical tests (see Fig.~\ref{RTOLATOLMAP}) have been
carried out to detect and avoid artificial numerical chaos arising from the
natural time discretisation of the applied numerical integration algorithms.
Our test determined an optimal choice in the absolute and relative error
tolerances required in the Gragg-Bulirsh-Stoer algorithm to detect real chaotic
dynamics. We have  calculated the MEGNO factor for several known irregular
satellites for the  prograde and retrograde cases. Our calculation suggests
that prograde  satellites are more chaotic in contrast to retrograde orbits 
(see Fig.~\ref{Y.DifferentSatellites}) though the detected chaos is very weak.
To  clarify this suggestion more detailed computations including also
additional  planetary perturbations along with Solar tides are necessary. Such 
calculations resulting in an estimate of the Lyapunov indicator are currently
in progress for the major irregular satellites.

It is important to note that every numerical tool capable of distinguishing
between quasi-periodic and chaotic dynamics has limitations with regards to
claiming quasi-periodicity considering only a limited period of time. This
certainly is also the case for the MEGNO technique. In Fig.~\ref{anankemegno}
we computed $Y(t)$ and $\langle Y\rangle(t)$ for the retrograde irregular
satellite Ananke. The plot shows that after 1 Myr the orbit of Ananke exhibits
a chaotic orbit with $\langle Y \rangle(1~\textnormal{Myr}) \approx 4.0$. In
the 60 kyr integrations the orbit of Ananke would possibly be interpreted as
quasi-periodic with $\langle Y\rangle(t)$ deviating only slightly from 2.0.

Considering 35100 orbits we calculated the MEGNO indicator on a large grid in 
$(a,I)$-space known to be occupied by observed irregular satellites
(Fig.~\ref{megnomaps}). The resulting map revealed several interesting
dynamical structures and we compared our results with previous studies
addressing the question of the  orbital stability of jovian test satellites. We
found good qualitative agreement between chaotic (quasi-periodic) and unstable
(stable) regions as was found previously \citep{carruba2002,nesvorny2003}. In
particular we confirm the asymmetry of the stable region  when comparing
prograde and retrograde satellite orbits \cite{nesvorny2003}.  Retrograde
satellite orbits have access to a larger volume of phase space  characterised
by orbit stability. This result is in contrast to the prograde satellites for
which chaotic orbits and its associated instability occurs (for $e_{0}=0.20$)
at $a \approx  0.13$~AU  $(0.37~\Rhill, 272~\Rjup)$ and onwards (cf.
Fig.~\ref{megnomaps}). A  small region of $(a,I)$-space at $a \approx 0.14$ AU
have been detected to  indicate  quasi-periodicity. In addition we detected the
presence of mean motion resonances of retrograde satellites with the Sun. The
location of  several high order mean motion resonances were determined and
compared with the present population of retrograde irregular satellite
families. We find  that the orbital elements of the members of the Pasiphae
family are largely scattered as opposed to the Carme group exhibiting to occupy
a more compact  region in $(a,I)$-space. We preliminary explain this excess in
scatter of the  osculating elements due to the close proximity to mean-motion 
resonances. This postulate will be subject to a separate study currently
ongoing addressing the question whether high-order (retrograde) mean motion
resonances are capable of  dispersing an initial compact group of satellite
members.

To support our results obtained from MEGNO we also calculated and compared two
initial conditions associated to two retrograde satellite orbits. The first
were chosen to be close to the 5:1 mean-motion resonance and the second
initial condition were chosen to be just  outside the location of this
resonance. We then searched for signs of chaoticity and quasi-periodicity to
validate the results obtained from calculating MEGNO. Applying a time-running
smoothing window on the osculating elements our  analysis of the single orbit
computations support the results obtained from MEGNO. Initial conditions
started in chaotic regions are associated to libration/circulation of resonant
angles. Quasi-periodic initial conditions show only circulating behaviour.

Motivated by the success of applying the MEGNO technique to the dynamics of
irregular satellites we plan to conduct a large parameter survey to identify 
further chaotic regions within the Hill sphere of Jupiter. In addition we plan 
to include giant planet perturbations and generate similar MEGNO maps of 
observed populations of irregular satellites in orbit around the remaining 
giant planets in the Solar System.

\section*{Acknowledgments}
TCH grately acknowledges K.~Go{\'z}dziewski for the introduction to the
MEGNO technique. Numerical simulations has been performed on the supercomputing
facility at the  University of Copenhagen (DCSC-KU) run on behalf of the Danish
Centre for Scientific Computing and the SFI/HEA Irish Centre for High-End
Computing  (ICHEC). TCH is thankful to Aake Nordlund for providing access to
the (DCSC-KU) computing facility. Astronomical research at the Armagh
Observatory is funded by the Northern Ireland Department of Culture, Arts and
Leisure (DCAL). TCH acknowledge David Asher, Matija {\'C}uk and Mark Bailey for 
fruitful discussions.

\bibliographystyle{mn2e}
\bibliography{IrrSatPaperMNRAS}

%%%%%%%%%%%%%%%%%%%%%%%%%%%%%%%%%%%%%%%%%%%%%%%%%%%%%%%%%%%%%%%%%%%%%%%%%%%%%%%%
\end{document}